\documentclass[11pt]{article}
\pdfoutput=1
\usepackage{graphicx}
\usepackage{amsmath, amsthm, amssymb}
\usepackage{amsfonts}
\usepackage{mathrsfs}
\usepackage[mathscr]{eucal}
\usepackage{subfigure}
\usepackage{natbib}

\usepackage[indent]{caption2}

\setcaptionwidth{100mm}
\usepackage[left=3.7cm,right=2.7cm,bottom=3cm]{geometry}

\usepackage[utf8]{inputenc}

\theoremstyle{definition}
\newtheorem{defi}{Definition}
\newtheorem{rem}{Remark}
\newtheorem{ex}{Example}

\newcommand{\supp}{\operatorname{supp}}
\newcommand{\dist}{\operatorname{dist}}

\reversemarginpar

\newcommand{\sign}{\operatorname{sign}}

\begin{document}
\large
\title{A framework for list representation, enabling list stabilization through incorporation of gene exchangeabilities}
\author{Charlotte Soneson and Magnus Fontes\\Centre for Mathematical Sciences, Lund University\\Box 118, SE-221 00 Lund, Sweden\\e-mail: \{lottas, fontes\}@maths.lth.se}
\date{\today}
\maketitle
\noindent

\begin{abstract}
{Analysis of multivariate data sets from e.g. microarray studies frequently results in lists of genes which are associated with some response of interest. The biological interpretation is often complicated by the statistical instability of the obtained gene lists with respect to sampling variations, which may partly be due to the functional redundancy among genes, implying that multiple genes can play exchangeable roles in the cell. In this paper we use the concept of exchangeability of random variables to model this functional redundancy and thereby account for the instability attributable to sampling variations. We present a flexible framework to incorporate the exchangeability into the representation of lists. The proposed framework supports straightforward robust comparison between any two lists. It can also be used to generate new, more stable gene rankings incorporating more information from the experimental data. Using a microarray data set from lung cancer patients we show that the proposed method provides more robust gene rankings than existing methods with respect to sampling variations, without compromising the biological significance.}
\end{abstract}

\section{Introduction}
Since the advent of the microarray technology, high-throughput experiments generating vast amounts of data have been ubiquitous in genetics, for studying e.g. genome-wide patterns of gene expression and copy number alterations. The output of univariate analysis of such high-throughput experiments is often a {\em gene list}, consisting of genes related to some response of interest (e.g. the discrimination between groups or a quantitative trait). The gene list can be ordered or unordered (i.e. ranking the genes by their association to the response or just listing all genes whose association exceeds some threshold) and it can consist of all studied genes or only a subset. The challenge is then to interpret the obtained list in a biological context to understand the underlying processes and generate biologically valid hypotheses. An inherent problem compromising the interpretability of the observed gene lists is that they are often highly unstable, both with regards to small changes in the underlying data set and with regards to changes in the ranking method \citep{Fortunel_etal_03,Irizarry_etal_05,Michiels_etal_05,EinDor_etal_06,Fan_etal_06,Boulesteix_Slawski_09,Abraham_etal_10}. This could be due to redundancy in the cell machinery, i.e. the existence of many genes having similar functions in the cell and thereby being exchangeable in a given experimental list. In this case, the observed gene list depends on the selection of samples in the data set. This means that the functional overlap between two lists may be substantial even though the actual gene overlap is very small. Other possible causes of the apparent instability are noisy measurements and the generally small sample sizes in this type of experiments \citep{EinDor_etal_06,He_Yu_10}. 

In this paper we propose a method for stabilization of observed gene rankings, using information extracted from the experimental data. We employ the concept of exchangeability of random variables to quantify the functional redundancy among the genes and we propose a general framework for incorporating exchangeability into the representation of gene lists. 
In this framework, each list is represented as a vector in $\mathbb{R}^M$ where $M$ is the number of genes in some universal set, typically the genes measured by a microarray chip. Each entry of the list vector quantifies the contribution to the list from one of the genes. 
This representation allows straightforward comparison of any two gene lists by means of any of the conventional measures of similarity or dissimilarity defined on $\mathbb{R}^M\times\mathbb{R}^M$. This is in contrast to previously proposed methods for list comparison, which are tailored to compare specific types of lists. We show that using the proposed method, we obtain gene rankings that are more robust than the original lists against sampling variations in the underlying data, without compromising the relevance to the response. 

\section{Related work}
The stabilization of gene rankings has attracted considerable interest during the last decade. Some authors have addressed the ranking method directly and proposed methods providing more robust and accurate ranking results and differential expression detection for the ``large $p$, small $N$'' situation which is standard in biomedical applications (e.g. \citet{Tusher_etal_01,Perelman_etal_07}). 

Another way to obtain more stable rankings is to combine the information from several different rankings (e.g. \citet{Rhodes_etal_02,Rhodes_etal_04,Breitling_etal_04,DeConde_etal_06,Hong_Breitling_08,Abeel_etal_10}). 
An overview of the most well-known aggregation methods is given by \citet{Boulesteix_Slawski_09}. The most straightforward method is to compute some univariate statistic for each gene from the set of rankings and re-order the genes by their value of this statistic. The statistic can be e.g. the mean of the positions for the gene \citep[e.g.][]{Jurman_etal_08}, a rank product of the positions \citep{Breitling_etal_04} or the fraction of the rankings where the gene is among the top-$k$ genes for some $k$ \cite[e.g.][]{Pepe_etal_03,Jurman_etal_08}. There are also more complex aggregation methods for extracting an optimal top-$k$ list based on e.g. Markov chains \citep{DeConde_etal_06}.

Comparison of gene lists is an essential part of many algorithms, e.g. for enrichment analysis of gene sets \citep[e.g.][]{Draghici_etal_03,Subramanian_etal_05,EinDor_etal_06,Ackermann_Strimmer_09} and assessment of the stability of gene rankings \citep[e.g.][]{Jurman_etal_08,Jurman_etal_10,Abraham_etal_10}. In general, each of the existing methods can only be used to compare specific types of lists, e.g. two ordered lists with the same number of genes, or one ordered list and one short unordered list. Appendix I provides a brief overview of the most well-known list comparison methods and show that many of them can be cast in the here proposed framework if its components are chosen suitably. 

\section{Methods}
\subsection{Exchangeability of random variables}
Consider a probability triple $\left(\Omega,\mathcal{F},P\right)$ and let $X_1,\ldots,X_m$ denote random variables on $\Omega$, taking values in some space $\mathbb{M}$. Given $X_1,\ldots,X_m$ we define the multivariate random variable $$X_{1}\times\ldots\times X_{m}:\Omega\rightarrow\underbrace{\mathbb{M}\times\ldots\times\mathbb{M}}_{m}$$by $X_{1}\times\ldots\times X_{m}(\omega)=(X_{1}(\omega),\ldots,X_{m}(\omega)).$ To each random variable $X_{1}\times\ldots\times X_{m}$ there is an associated measure $Pr_{X_{1}\times\ldots\times X_{m}}$ defined by $$Pr_{X_{1}\times\ldots\times X_{m}}(A)=P\left(\left\{\omega\in\Omega;\,\,X_{1}\times\ldots\times X_{m}(\omega)\in A\right\}\right)$$ for all measurable subsets $A\subseteq\mathbb{M}\times\ldots\times\mathbb{M}$.  

Conventionally, a finite sequence $(X_1,\ldots,X_m)$ of random variables is called {\em exchangeable} if their joint distribution is invariant under permutation of $X_1,\ldots,X_m$, i.e. if $$Pr_{X_1\times\ldots\times X_m}=Pr_{X_{\pi(1)}\times\ldots\times X_{\pi(m)}}$$ for each $\pi\in S_m$ (the group of permutations of $\{1,\ldots,m\}$). This means that from a statistical point of view the order of the variables in the product is completely irrelevant. 
From this definition, it is clear that any sequence of independent and identically distributed (i.i.d.) random variables is exchangeable, but the reverse implication is false. For overviews on exchangeability, see e.g. \citet{Kingman_78,Aldous_85}. The definition of exchangeability given above is rather strong, and we introduce a much weaker notion of exchangeability as follows:
\begin{defi}\label{def:exchangeable} The finite sequence of random variables $\left( X_1,\ldots, X_m\right)$ is {\em weakly exchangeable} if the null sets of $Pr_{ X_1\times\ldots\times X_m}$ are invariant under permutations, i.e. $$Pr_{X_{\pi(1)}\times\ldots\times X_{\pi(m)}}<<Pr_{X_{\tau(1)}\times\ldots\times X_{\tau(m)}}$$ for all $\pi,\tau\in S_m$. Here $\mu<<\nu$ denotes that the positive measure $\mu$ is absolutely continuous with respect to the positive measure $\nu$.\end{defi}
It is clear that a finite sequence of random variables $\left( X_1,\ldots, X_m\right)$ that is exchangeable is weakly exchangeable, but that the opposite implication is false in general.

\subsection{Measures of exchangeability}
In this section we will discuss some ways to quantify the degree of exchangeability for a sequence of random variables.
\begin{defi}Given a finite sequence of random variables $\left( X_1,\ldots, X_m\right)$, the {\em total exchangeability variation} is given by $$P_{ X_1\times\ldots\times X_{m}}^{Var}:=\frac{1}{|S_m|-1}\sum_{\pi\in S_m}\left\|Pr_{X_{\pi(1)}\times\ldots\times X_{\pi(m)}}-\frac{1}{|S_m|}\sum_{\tau\in S_m}Pr_{X_{\tau(1)}\times\ldots\times X_{\tau(m)}}\right\|.$$ Here, $\|\mu\|$ denotes the total variation of the (real-valued) measure $\mu$.\end{defi}
We note that $P_{ X_1\times\ldots\times X_ m}^{Var}=0$ iff the sequence $( X_1,\ldots, X_m)$ is exchangeable. 

We now turn to a discrete probability space $\left(\Omega,\mathcal{F},P\right)$, where $\Omega$ is a finite set, $\mathcal{F}=2^\Omega$ is the $\sigma$-algebra consisting of all events and $P:\mathcal{F}\rightarrow[0,1]$ is a probability measure. 
We let $X_1,\ldots,X_m$ be random variables on $\Omega$ taking values in $\mathbb{M}:=\{1,\ldots,M\}$. The {\em support} of the random variable $X_{1}\times\ldots\times X_{m}$ is defined by $$\supp X_{1}\times\ldots\times X_{m}:=\left\{\left(q_1,\ldots,q_m\right)\in\mathbb{M}\times\ldots\times\mathbb{M};\,Pr_{X_{1}\times\ldots\times X_{m}}\left(\left\{\left(q_1,\ldots,q_m\right)\right\}\right)>0\right\}.$$For finite sequences of discrete random variables, Definition~\ref{def:exchangeable} implies that $(X_1,\ldots,X_m)$ is weakly exchangeable iff $$\supp\left(X_1\times\ldots\times X_m\right)=\supp\left(X_{\pi(1)}\times\ldots\times X_{\pi(m)}\right)$$for all $\pi\in S_m$. Therefore, to quantify the degree of weak exchangeability for a sequence of discrete random variables we will compare the support of the joint distributions. 

Let {$\rho:(\mathbb{M}\times\ldots\times\mathbb{M})\times(\mathbb{M}\times\ldots\times\mathbb{M})\rightarrow\mathbb{R}$} be a metric and define the distance between two sets $A,B\subseteq\mathbb{M}\times\ldots\times\mathbb{M}$ by $$\dist_\rho(A,B):=\min_{a\in A,b\in B}\rho(a,b).$$Furthermore, define the Hausdorff distance between the two sets by $$HD_\rho(A,B):=\max\left(\sup_{a\in A}\dist_\rho(\{a\},B),\sup_{b\in B}\dist_\rho(\{b\},A)\right).$$
\begin{defi}Given a finite sequence of discrete random variables $(X_1,\ldots,X_m)$, the {\em maximal exchangeability distance} is given by $$ED^{max}_{X_1\times\ldots\times X_m}=\frac{\sum_{\pi\in S_m}\sum_{\tau\in S_m}HD_\rho\left(\supp X_{\pi(1)}\times\ldots\times X_{\pi(m)},\supp X_{\tau(1)}\times\ldots\times X_{\tau(m)}\right)}{\rho\left(1_m,M_m\right)|S_m|(|S_m|-1)}$$and the {\em mean exchangeability distance} is given by \begin{align*}ED&^{mean}_{X_1\times\ldots\times X_m}\\&=\frac{\sum_{\pi\in S_m}\sum_{\tau\in S_m}\mathbb{E}_{X_{\pi(1)}\times\ldots\times X_{\pi(m)}}\left[\dist_\rho\left(X_{\pi(1)}\times\ldots\times X_{\pi(m)},\supp X_{\tau(1)}\times\ldots\times X_{\tau(m)}\right)\right]}{\rho\left(1_m,M_m\right)|S_m|(|S_m|-1)}.\end{align*}Here, $1_m=\left(1,\ldots,1\right)$ and $M_m=\left(M,\ldots,M\right)$.\end{defi}
Clearly, $ED^{max}_{X_1\times\ldots\times X_m}=ED^{max}_{X_{\pi(1)}\times\ldots\times X_{\pi(m)}}$ for any $\pi\in S_m$, and $ED^{max}_{X_1\times\ldots\times X_m}=0$ iff $\left(X_1,\ldots,X_m\right)$ is weakly exchangeable, and the same is true for $ED^{mean}_{X_1\times\ldots\times X_m}$. 

In the rest of this paper we will mainly consider exchangeability of pairs of discrete random variables with values in $\mathbb{M}=\{1,\ldots,M\}$. In this special case, since $X_1\times X_2$ is the reflection of $X_2\times X_1$ with respect to the line $\{(x,y)\in\mathbb{R}^2;\,y=x\}$, we get \begin{align}P^{Var}_{X_1\times X_2}&=\|Pr_{X_1\times X_2}-Pr_{X_2\times X_1}\|\label{eq:totalvariationpair}\\ED^{max}_{X_1\times X_2}&=\frac{HD_\rho\left(\supp X_1\times X_2,\supp X_2\times X_1\right)}{\rho\left((1,1),(M,M)\right)}\label{eq:maxdistpair}\\ED^{mean}_{X_1\times X_2}&=\frac{\mathbb{E}_{X_1\times X_2}\left[\dist_\rho\left(X_1\times X_2,\supp X_2\times X_1\right)\right]}{\rho\left((1,1),(M,M)\right)}\label{eq:meandistpair}.\end{align}

\subsection{The exchangeability plot}
To illustrate the degree of weak exchangeability of a pair of discrete random variables visually we propose the {\em exchangeability plot}. The exchangeability plot for the random variables $ X_1$ and $ X_2$ is obtained by depicting both $\supp X_1\times X_2$ and $\supp X_2\times X_1$ in the same figure. A pair of random variables $(X_1,X_2)$ is weakly exchangeable iff the two sets overlap completely. Figure~\ref{fig:exchplots} shows two exchangeability plots. The pair of variables in the left panel is weakly exchangeable, while the pair of variables in the right panel is not. Given samples of $X_1\times X_2$ and $X_2\times X_1$ we can also define a {\em sample exchangeability plot}, depicting the observed supports. Further details are provided in Appendix A.

\subsection{The exchangeability of genes}
In this section we use the terminology developed in the previous sections to define the exchangeability of a set of genes with respect to a specific experiment. We will also define another measure of exchangeability which is specifically adapted to the study of ranked gene lists. 

We assume that we are given a universal set of $M$ genes, denoted $\mathcal{G}=\{g_1,\ldots,g_M\}$. We use the word {\em experiment} to denote a pair consisting of a population (e.g. cancer patients and healthy control subjects) and a variable ranking method (e.g. a t-test contrasting the two groups in the population). The sample space $\Omega$ consists of all possible rankings of the $M$ genes, and the random variables $X_1,\ldots, X_M:\Omega\rightarrow\{1,\ldots,M\}$ represent the ranking positions of the genes in $\mathcal{G}$. A finite set of genes $\{g_{i_1},\ldots,g_{i_m}\}$ is said to be (weakly) exchangeable iff the corresponding sequence of random variables $(X_{i_1},\ldots,X_{i_m})$ is (weakly) exchangeable. Intuitively, a set of genes is exchangeable iff their positions in the variable ranking can be interchanged without changing the biological interpretation of the ranking. 

Of course, we do not know the measures $Pr_{X_i}$ for the variables in practice, so these have to be estimated. In this paper we use subsampling to generate a collection of $B$ data sets for which we compute gene rankings. From the $B$ gene rankings, we construct a {\em position vector} $S_i$ for each gene $g_i$ by collecting all positions of the gene in the $B$ rankings. The elements of a position vector $S_i$ are then samples of the random variable $X_i$. Combining two position vectors $S_i$ and $S_j$ gives samples of the variables $X_i\times X_j$ and $X_j\times X_i$ which can be used to obtain estimates of $P^{Var}_{X_i\times X_j}$, $ED^{max}_{X_i\times X_j}$ and $ED^{mean}_{X_i\times X_j}$. This estimation is discussed further in Appendix A.

We now introduce another measure of exchangeability which is especially adapted to the study of ranked gene lists. The rationale behind this measure is that we may not want two genes to obtain a small exchangeability distance if they always appear in the same order in the rankings. For example, a gene which is always ranked first is not highly exchangeable with a gene that is always ranked second, since the first gene is clearly more strongly related to the response than the second. The new measure penalizes such situations in the computation of the exchangeability distance. For a pair of random variables $(X_1,X_2)$ we define a new set-valued random variable on $\Omega$ by 
$$R\left(X_1\times X_2\right)\left(\omega\right):=\{(x,y)\in\mathbb{M}\times\mathbb{M};\,\,\sign\left(x-y\right)=\sign\left(X_1\left(\omega\right)-X_2\left(\omega\right)\right)\}.$$

\begin{defi}The {\em one-sided mean exchangeability distance} for a pair of discrete random variables $(X_1,X_2)$ is defined by $$oED^{mean}_{X_1\times X_2}:=\frac{\mathbb{E}_{X_1\times X_2}\left[\dist_\rho\left(X_1\times X_2,\supp X_2\times X_1\bigcap R(X_1\times X_2)\right)\right]}{\rho\left(\left(1,2\right),\left(M-1,M\right)\right)}$$if $\supp X_2\times X_1\cap R(X_1\times X_2)(\omega)\neq\emptyset$ for all $\omega\in\Omega$ with $Pr_{X_1\times X_2}(X_1\times X_2(\omega))>0$, 
and $oED^{mean}_{X_1\times X_2}=1$ otherwise.\end{defi}

Details on the estimation of the one-sided mean exchangeability distance are provided in Appendix A. It is also possible to define a one-sided variant of the maximal exchangeability distance ($oED^{max}_{X_i\times X_j}$) in an analogous manner. 

We note that due to the normalization factors introduced in the estimates of weak exchangeability, all measures introduced above attain only values in $[0,1]$. This allows us to define similarity measures ({\em exchangeability scores}) for pairs of genes as follows: $$\begin{array}{lllll}PS^{Var}_{X_i\times X_j}=1-P^{Var}_{X_i\times X_j},&&ES^{mean}_{X_i\times X_j}=1-ED^{mean}_{X_i\times X_j},&&ES^{max}_{X_i\times X_j}=1-ED^{max}_{X_i\times X_j},\\oES^{mean}_{X_i\times X_j}=1-oED^{mean}_{X_i\times X_j},&&oES^{max}_{X_i\times X_j}=1-oED^{max}_{X_i\times X_j}.\end{array}$$
Finally, we define normalized values of the exchangeability scores by comparing them to the corresponding values for pairs of random variables with some pre-specified distribution representing a null hypothesis of no association. In this paper, the main focus is on weak exchangeability of pairs of discrete random variables, in which case it is natural to compare to a random variable $Y_1\times Y_2$ uniformly distributed on a set $S\subseteq\mathbb{M}\times\mathbb{M}$ with cardinality equal to that of $\supp X_1\times X_2$. We show only the normalization for $oES^{mean}_{X_i\times X_j}$, the other scores can be normalized analogously.
\begin{defi}The {\em normalized one-sided mean exchangeability score} for a pair of discrete random variables $(X_1,X_2)$ is defined by \begin{equation}\label{form:normalizedexch}noES^{mean}_{X_1\times X_2}=\left(\frac{oES^{mean}_{X_1\times X_2}-oES^{mean}_{Y_1\times Y_2}}{1-oES^{mean}_{Y_1\times Y_2}}\right)_+\end{equation}where $Y_1\times Y_2$ is a random variable uniformly distributed on a set $S\subseteq\mathbb{M}\times\mathbb{M}$ with $|S|=|\supp X_1\times X_2|$, and $(a)_+=\max(a,0)$. \end{defi}

\begin{figure}[htb]
\begin{center}
\subfigure{\resizebox{55mm}{55mm}{\label{exchplots:a}\includegraphics{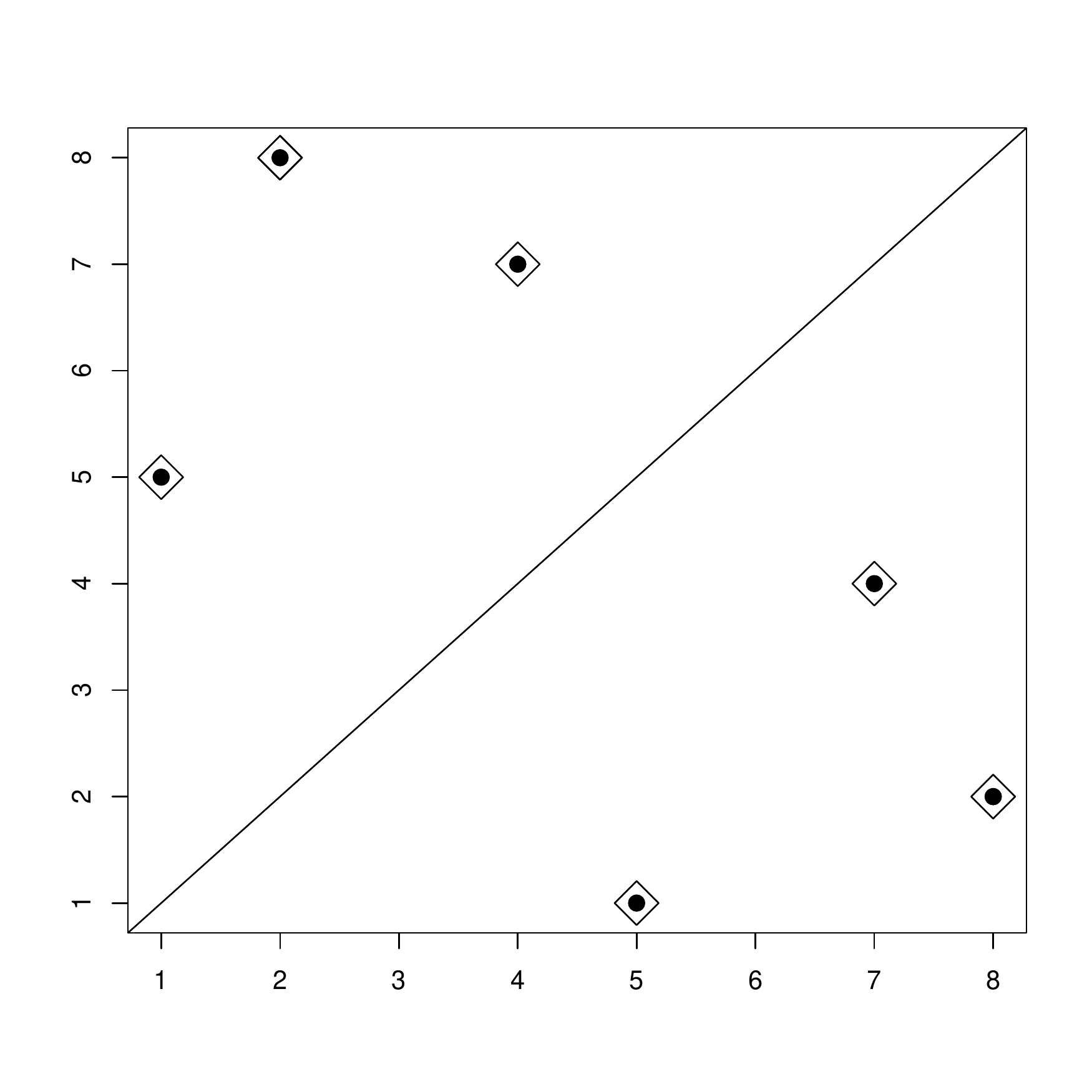}}}
\subfigure{\resizebox{55mm}{55mm}{\label{exchplots:b}\includegraphics{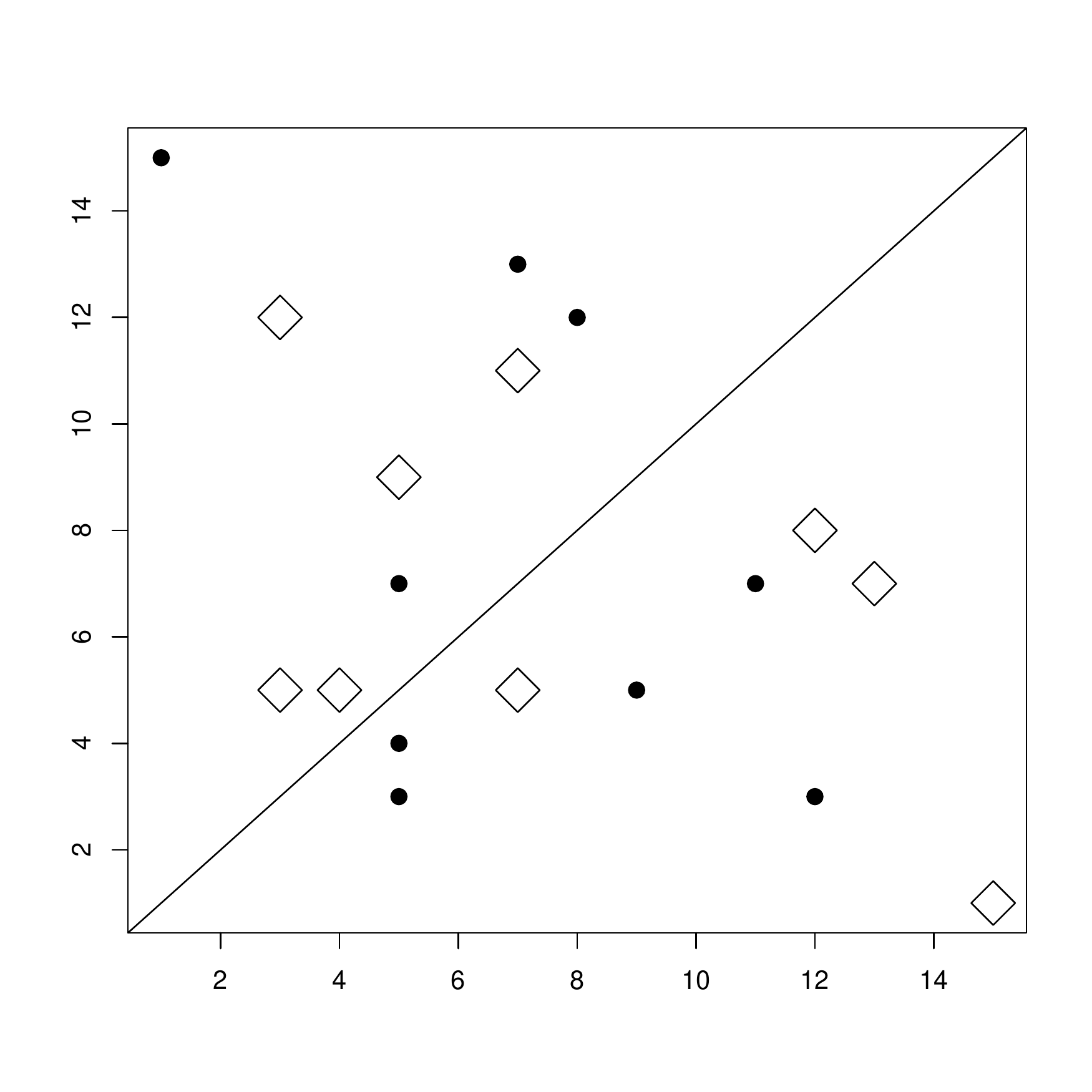}}}
\caption{Exchangeability plots for two pairs of random variables. The pair in the left panel is weakly exchangeable, while the pair in the right panel is not. }
\label{fig:exchplots}
\end{center}
\end{figure}

We note that the measures of exchangeability depend on the number of genes in the ranking ($M$). For two genes having the exchangeability plot shown in the left panel of Figure~\ref{fig:exchplots} we get ${noES}^{mean}_{X_1\times X_2}=1.0$ irrespective of the number of genes since in this case, the distance between any value of $X_1\times X_2$ and $\supp X_2\times X_1$ is zero. For the exchangeability plot in the right panel of Figure~\ref{fig:exchplots} we obtain ${noES}^{mean}_{X_1\times X_2}=0.17$ if $M=15$ and ${noES}^{mean}_{X_1\times X_2}=0.99$ if $M=1,000$. In Appendix H we compute exchangeability matrices for some synthetic example data sets.

\section{A general framework for list representation and comparison}\label{sec:framework}
In this section, we present a general framework for list representation and comparison. The lists are represented as vectors in $\mathbb{R}^M$, where the entry in position $i$ gives the contribution of gene $g_i$. The vector representation allows us to compare both ranked and unranked gene lists within the same framework, using one of the many similarity or dissimilarity measures available to compare vectors in $\mathbb{R}^M$. This is an advantage compared to existing methods for list comparison, which are specifically designed to compare certain types of lists. Our framework also provides a way to determine which genes are most important for explaining the similarity between two lists. 
Assume for example that some measure based on the scalar product in $\mathbb{R}^M$ is used to measure the similarity between two vectors $x$ and $y$. Then the value of $x_iy_i$ is a measure of the influence of the $i$'th variable on the similarity between the two lists (see Appendix G). 
Finally, ordering the genes by their weights in the vector gives a new ranking of the genes. 

As above, we have a universal set of $M$ genes, $\mathcal{G}=\{g_1,\ldots,g_M\}$, where the genes are indexed in a fixed (but otherwise arbitrary) fashion. The universal set can be e.g. all genes on a microarray chip. An ordered (unordered) {\em gene list} is then an ordered (unordered) subset of the universal set. By defining a function $$f:(positions,exchangeabilities,reliability,...)\mapsto l_\ell\in\mathbb{R}^M$$ we use information about various characteristics of the given list to create a vector representation. 

\subsection{General idea}
Let $\ell\subseteq\mathcal{G}$ denote a list. If $\ell$ is ordered and if gene $g_i$ is contained in $\ell$, we denote its position by $\pi_\ell(i)$. If $g_i\not\in\ell$, we define $\pi_\ell(i)=0$. For an unordered list, we let $\pi_\ell(i):=\chi_\ell(g_i)$, where $\chi_\ell$ is the characteristic function of the set $\ell$. 
Given a list $\ell$ we define a corresponding {\em list matrix} $G_\ell$ as the product of three basic $M\times M$ matrices; $$G_\ell:=A_\ell V_\ell W_\ell.$$The three basic matrices are designed to represent different characteristics of $\ell$. We call $A_\ell$ the {\em position matrix}, $V_\ell$ the {\em exchangeability matrix} and $W_\ell$ the {\em global weight matrix}. From the list matrix we form a {\em list vector} $l_\ell:=((l_\ell)_1,\ldots,(l_\ell)_M),$ by letting $(l_\ell)_i:=h((G_\ell)_i)$ where $(G_\ell)_i$ denotes the $i$'th column of $G_\ell$ and $h:\mathbb{R}^M\rightarrow\mathbb{R}$ is a summarization function, e.g. a norm. 
The list vector will be used as the vector representation of the list. Once all lists of interest are represented by vectors in $\mathbb{R}^M$, we can define the similarity between them e.g. as the cosine of the angle between the corresponding list vectors, i.e. $$s(\ell_1,\ell_2)=\frac{l_{\ell_1}\cdot l_{\ell_2}}{\|l_{\ell_1}\|_2\|l_{\ell_2}\|_2},$$where $\cdot$ denotes the inner product in $\mathbb{R}^M$, and we can obtain a dissimilarity coefficient as \begin{equation}\label{eq:distmeas}d(\ell_1,\ell_2)=1-s(\ell_1,\ell_2).\end{equation} 

Choosing $A_\ell, V_\ell, W_\ell, h$ and the (dis)similarity coefficient on $\mathbb{R}^M$ suitably, most methods currently available for list comparison fit into this general framework. In Appendix I we show how this can be done for a collection of well-known methods.

\subsection{The position matrix $A_\ell$}
The position matrix $A_\ell$ is defined as a diagonal matrix that contains information about the type of list (ordered or unordered) and the positions of the genes within the list. We define the diagonal element $(A_\ell)_{ii}$ (the position value of gene $g_i$) via a monotone transformation of the ranking statistic of the gene.
This means that the diagonal elements corresponding to genes in the top of the list $\ell$ are high, while the genes further down in the list obtain lower values. All genes not in the list are given position value zero. We note that in some cases, other choices of position values may be better suited for unordered lists, where it may be desirable to give the genes different weights, e.g. depending on some external criterion, even though there is no specified ordering.

\subsection{The exchangeability matrix $V_\ell$}
The exchangeability matrix $V_\ell$ carries information about the exchangeability between the genes in $\mathcal{G}$ in the specific experiment giving the list $\ell$. In most examples in this paper, we define the entry $(V_\ell)_{ij}$ to be the estimated normalized one-sided mean exchangeability score of $g_i$ and $g_j$ (i.e. $\widehat{noES}^{mean}_{X_i\times X_j}$), so the diagonal elements are always 1.
If $V_\ell$ is diagonal, i.e. $V_\ell=I_M$, then the only non-zero elements in the list vector are those corresponding to genes that are actually contained in $\ell$ and consequently only the genes that are present in the list affect its vector representation. However, if $V_\ell$ is not diagonal, there is a possibility that the vector representation of the list is extended, i.e. that it contains non-zero entries for genes which are not themselves present in the list, but are exchangeable with some gene in the list. The (absolute) weight of a gene in the list can also be increased if it is highly exchangeable with a gene with a higher (absolute) position value, since the high exchangeability indicates that the genes could as well have switched positions without changing the interpretation of the list. 

We note that the general framework for list representation supports any matrix of gene similarities in the place of $V_\ell$. For example, $V_\ell$ could be defined from some kind of expert biological knowledge, e.g. concerning which genes are related to the same biological function. This could be used for example when comparing lists from different experiments to each other. 
Yet another option is to use the positive part of the correlation matrix in place of $V_\ell$, since a high correlation between the expression levels of two genes is often considered to indicate similar biological functions of the genes. 

\subsection{The global weight matrix $W_\ell$}
The global weight matrix $W_\ell$ is a diagonal matrix that permits weighting the influence of the genes differently, depending on some informativeness or reliability estimate. For example, we may wish to downweight the influence of a gene that has a high probability to be present in an arbitrarily chosen list, since this gene is unspecific and may not give much relevant information about the similarity between a pair of lists. 

\section{Applications}
\subsection{Data sets}
To illustrate the proposed methods, we use three microarray data sets, which were downloaded from \textit{http://www.broadinstitute.org/gsea/datasets.jsp}. These data sets have already been pre-processed by replacing the original probe set IDs with gene symbols and summarizing all probe sets mapping to the same gene by the largest value for each sample. The two lung cancer data sets were re-analyzed by \citet{Subramanian_etal_05}. 
\begin{itemize}\item{\bf Boston lung cancer data} \citep{Bhattacharjee_etal_01}. This data set contains expression measurements of 5,217 genes in 62 lung cancer patients, classified according to outcome (good or poor) with 31 observations in each group. 
\item{\bf Michigan lung cancer data} \citep{Beer_etal_02}. This data set contains expression measurements of the same 5,217 genes as in the Boston lung cancer data, for 86 lung cancer patients (24 with poor outcome and 62 with good outcome).
\item{\bf Diabetes data} \citep{Mootha_etal_03}. This data set contains expression measurements of 15,056 genes in 17 diabetic patients and 17 control subjects. 
\end{itemize}

\subsection{Stabilization of ranked gene lists}\label{sec:stability}
The main objective for introducing exchangeabilities into the list representation is to increase the robustness of the resulting gene list. In this section we evaluate different aspects of the stability of the extended gene lists by comparing with non-extended lists, lists extended using correlations and lists generated by aggregation. 

\subsubsection{Stability of top-$k$ gene lists}\label{sec:rankingstability}
In many cases, only the top-ranked genes from an experiment are studied further, which stresses the importance of obtaining a robust and informative set of top-ranked genes. To study the usefulness of exchangeability stabilization for this purpose we apply the following steps to the Boston lung cancer data. 
\begin{enumerate}
\item We generate 10 modified data sets by bootstrapping samples from the original data set, taking the class labels into account.
\item\label{step:createrankings} For each modified data set, we rank the genes using five different methods:
\begin{enumerate}
\item Ranking the genes according to their signal-to-noise ratio (SNR) when comparing the two patient groups. Genes positively associated with good outcome are placed in the top. Here, $$SNR(i) = \frac{m_i(good)-m_i(poor)}{\sigma_i(good)+\sigma_i(poor)}$$where $m_i(poor)$ and $\sigma_i(poor)$ denote the mean value and standard deviation of gene $i$ in the patients with poor outcome, and $m_i(good)$ and $\sigma_i(good)$ are the corresponding values in the patients with good outcome. Ranking the genes by their SNR values gives the {\bf non-extended lists}. In general, lists generated in this or similar ways are those that are used for interpretation and biological conclusions. 

\item\label{step:extended} Computing the extended list vector as described in Section~\ref{sec:framework} and ranking the variables according to their contribution to the list vector. The position matrix is derived from the SNR-based ranking of the genes, by letting $$(A_\ell)_{ii}=\left\{\begin{array}{cl}\frac{b^2}{(\pi_\ell(i)-1)^2+b^2}&\textrm{if }SNR(i)\geq 0\\-\frac{b^2}{(M-\pi_\ell(i))^2+b^2}&\textrm{if }SNR(i)<0,\end{array}\right.$$with $b^2=350$ and $M=5,217$. We comment on the selection of this function and the parameter values in Appendix B. The position vectors are computed by subsampling the original data set $B=20$ times (each time keeping 2/3 of the samples from each group) and ranking the variables by their SNR values. From the position vectors we then compute the normalized one-sided mean exchangeability scores $\widehat{noES}^{mean}_{X_i\times X_j}$ for all gene pairs to create the exchangeability matrix. We take the global weight matrix $W_\ell=I_M$. To create the list vector from the list matrix, we define 
the $i$'th entry of the list vector as the element with the largest magnitude in the $i$'th column of the list matrix $G_\ell$. This means that a gene which is strongly exchangeable with a gene with a highly negative position value can be moved downwards in the list, so the two extreme ends are treated symmetrically. In the final ranking, genes with a highly positive contribution are placed in the top and genes with a highly negative contribution are placed in the bottom. This gives the {\bf extended lists}.

\item Subsampling the modified data set 100 times, and each time deducing a ranking from the SNR values as above. The final ranking is then obtained as an aggregate ranking, by computing the median position of each gene in the 100 subsample rankings. The gene with the lowest median position is placed in the top. This procedure gives the {\bf median aggregated lists}. We also create aggregated lists by computing the product of the ranks of each gene in the 100 subsample rankings and ordering the genes by increasing value of the rank product. This gives the {\bf rank product aggregated lists}. 

\item Computing the extended list vector as described in step~\ref{step:extended}, but using the positive part of the correlation matrix of the original data set as the exchangeability matrix. The genes are ordered by decreasing contribution to the resulting list vector. This gives the {\bf correlation extended lists}.  
\end{enumerate}

\item The correspondence between the lists from different bootstrap data sets are visualized through concordance plots. For each of the five ranking methods described in step~\ref{step:createrankings}, let $f_k$ denote the number of genes that are among the top-$k$ in the resulting lists from all bootstrapped data sets. The concordance plot depicts $f_k$ as a function of $k$ for $k\in\{1,\ldots,5217\}$. If the lists are highly reproducible with respect to sampling variation in the underlying data, we get $f_k\approx k$ for all $k$. We also construct concordance plots for the reversed lists, i.e. letting $f_k$ be the number of genes that are among the bottom-$k$ in all lists. As another measure of the stability of the gene rankings, we compute the mean overlap between the top-30 and bottom-30 genes from each pair of bootstrapped data sets for each of the five ranking methods defined in step~\ref{step:createrankings}. The resulting figures are given in Appendix E.
\end{enumerate}

The top row in Figure~\ref{fig:concordance} shows concordance plots for the gene lists obtained by the five ranking methods. It is clear that the exchangeability-extended lists are more stable than the lists obtained by the other methods with respect to sampling variations in the underlying data set. Notably, the correlation-extended lists are less stable than the exchangeability-extended lists, indicating that the correlations in this case do not capture the relevant characteristics of the data. The bottom row in Figure~\ref{fig:concordance} shows corresponding concordance plots for gene lists extracted from a data set where the sample labels have been randomly permuted. These figures show that the stability of the extended lists that was noted in the top row is clearly dependent on that the gene lists actually share some information. Hence, the stability is not due to spurious features unrelated to the discrimination between patients with good and poor outcome. For this data set, the correlation between the expression values for a pair of genes has little to do with the estimated exchangeability of the genes (see Appendix D). 

\subsubsection{Stability of distances between ranked lists}
Next, we study the stability of the distance between ranked lists from the three different data sets. 
First, the data sets are adjusted to contain the same genes, which leaves 5,149 genes. We then compute the exchangeability matrix for each data set (normalized one-sided mean exchangeability scores, position vectors obtained by subsampling $B=20$ times). For each data set, we construct 10 modified data sets by bootstrapping, taking the class labels into account, and compute extended and non-extended list vectors for each bootstrapped data set. The pairwise distances between all extended (or non-extended, respectively) list vectors are computed using (\ref{eq:distmeas}) and we study the variation of the distances from each comparison. For list vectors from different data sets, the aim is to obtain a robust value of the distance. For list vectors from the same data set the distance should also be close to zero. 

Figure~\ref{fig:distancestabilityrankedlists} shows histograms of the computed distances for each comparison. It is clear that the lists from the same data set are much more similar after extension than before (the distances are much closer to zero). Comparing the two lung cancer data sets, the extended list vectors are more similar and the distance estimates are more stable than without extension. This suggests that the extended list vectors incorporate information which is shared between the two data sets and that is missed if we only study the top genes. For the comparisons between the lung cancer data sets and the diabetes data set, the non-extended list vectors are almost orthogonal (implying a dissimilarity around 1), which indicates that the top genes from the data sets are completely different. The dissimilarities are close to 1 also after extension, which suggests that there is much less shared functional information between a lung cancer data set and the diabetes data set than between the two lung cancer data sets. In Appendix F we give the corresponding histograms for rankings obtained from data sets were the class labels are permuted independently in each bootstrap round.

\subsection{Informativeness of ranked gene lists}
Although stability of gene rankings is an important and desirable property, it is not the only thing that is of interest. For example, if we define a ranking method which always assigns a gene the same pre-defined position, the ranking would be extremely stable but most likely useless. We therefore study the informativeness of the rankings obtained as described above by examining the ability of the top-ranked genes in each list to discriminate between the two patient groups in the Boston lung cancer data. We use ten-fold cross-validation to assess the performance of the classifiers. For each training/test set split we compute the five rankings as described in Section~\ref{sec:rankingstability} for the training set, and extract the top- and bottom-$k$ genes from each ranking. The expression levels for these genes are centered and standardized based on their mean value and standard deviation in the training set. The standardized expression levels of the selected genes are then used as features in a centroid classifier \citep{Scholkopf_Smola_02} which is used to classify the remaining (test) samples. The reported classification accuracy is the mean area under the receiver operating characteristic curve (AUC) across the 10 training/test set splits. 
Table~\ref{tab:classificationability} shows the estimated classification accuracy for the top and bottom genes from the five rankings as well as the mean classification accuracy for top and bottom genes from 20 random rankings. Note that the top-ranked gene is always the same for the extended and non-extended rankings. The classification ability of the top-ranked genes in the extended list vectors is considerably higher than for the other methods, indicating that the increased stability observed in the previous section does not come at the expense of decreased biological significance. \begin{table}
\caption{Classification accuracy (mean AUC across 10 training/test set splits) for the union of the top- and bottom-$k$ lists obtained from different rankings of the genes in the Boston Lung Cancer data with respect to their association with the discrimination between patients with good and poor outcome. The best performing method for each $k$ is highlighted in bold. }
\centering
\begin{tabular}{lcccc}

&$k=1$&$k=10$&$k=30$&$k=100$\\
\hline
Extended&0.344&{\bf 0.774}&{\bf 0.837}&{\bf 0.832}\\
Non-extended&0.344&0.583&0.606&0.566\\
Median aggregated&0.410&0.565&0.617&0.566\\
Rank product&0.400&0.538&0.606&0.566\\
Correlation extended&0.344&0.594&0.572&0.594\\
Random&{\bf 0.464}&0.489&0.473&0.478\\
\end{tabular}
\label{tab:classificationability}
\end{table}

\section{Discussion}
Univariate analysis of multivariate genetic data sets usually results in a ranking of the variables according to some criterion. This ranking is then interpreted to gain biological knowledge and understanding. However, it has been noted that the variable rankings are often highly unstable with respect to small changes in the underlying data set or the method used to obtain the ranking and therefore, methods for stabilizing the variable ranking and allowing more robust comparison to other lists are much needed. In this paper we have presented a general framework for robust representation and comparison of variable lists. The framework encompasses both ordered and unordered lists, which can therefore be compared on similar terms. Having a robust measure of similarity between any pair of variable lists can furthermore enable visualization through e.g. multidimensional scaling to obtain a low-dimensional visual representation of large collections of lists. We have shown that the extended variable lists are more stable than the original variable lists from an experiment, and also more stable than lists obtained by aggregation of several lists from subsampled data sets. These results suggest that the exchangeability concept for random variables may be a suitable tool for quantifying the functional redundancy among genes and incorporating this information into the list representation. 

The results from the proposed method can be used in different ways. Given the vector representation of the gene lists there are many natural choices of similarity and dissimilarity measures that can be applied to compare lists. By ranking the genes according to their contribution to the list vector we also obtain a new gene ranking which may be used to obtain more robust results with other methods, such as e.g. Gene Set Enrichment Analysis (GSEA) \citep{Subramanian_etal_05} to study the enrichment of gene sets among the genes most highly related to a response.

\bibliographystyle{natbib}
\bibliography{ListComparisonBibliography}

\clearpage

\begin{figure}[htb]
\begin{center}
\subfigure{\resizebox{55mm}{!}{\label{concordance:a}\includegraphics{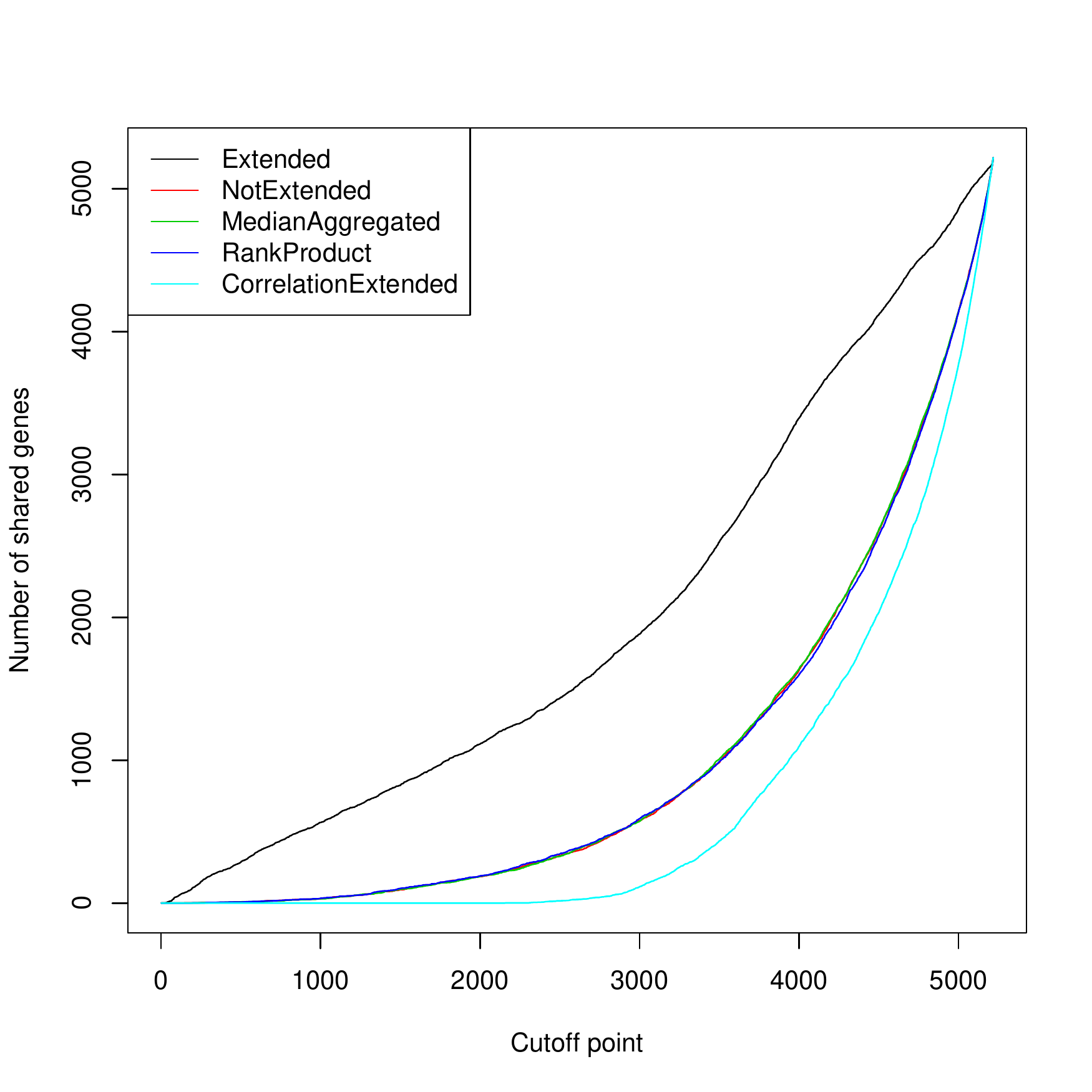}}}
\subfigure{\resizebox{55mm}{!}{\label{concordance:b}\includegraphics{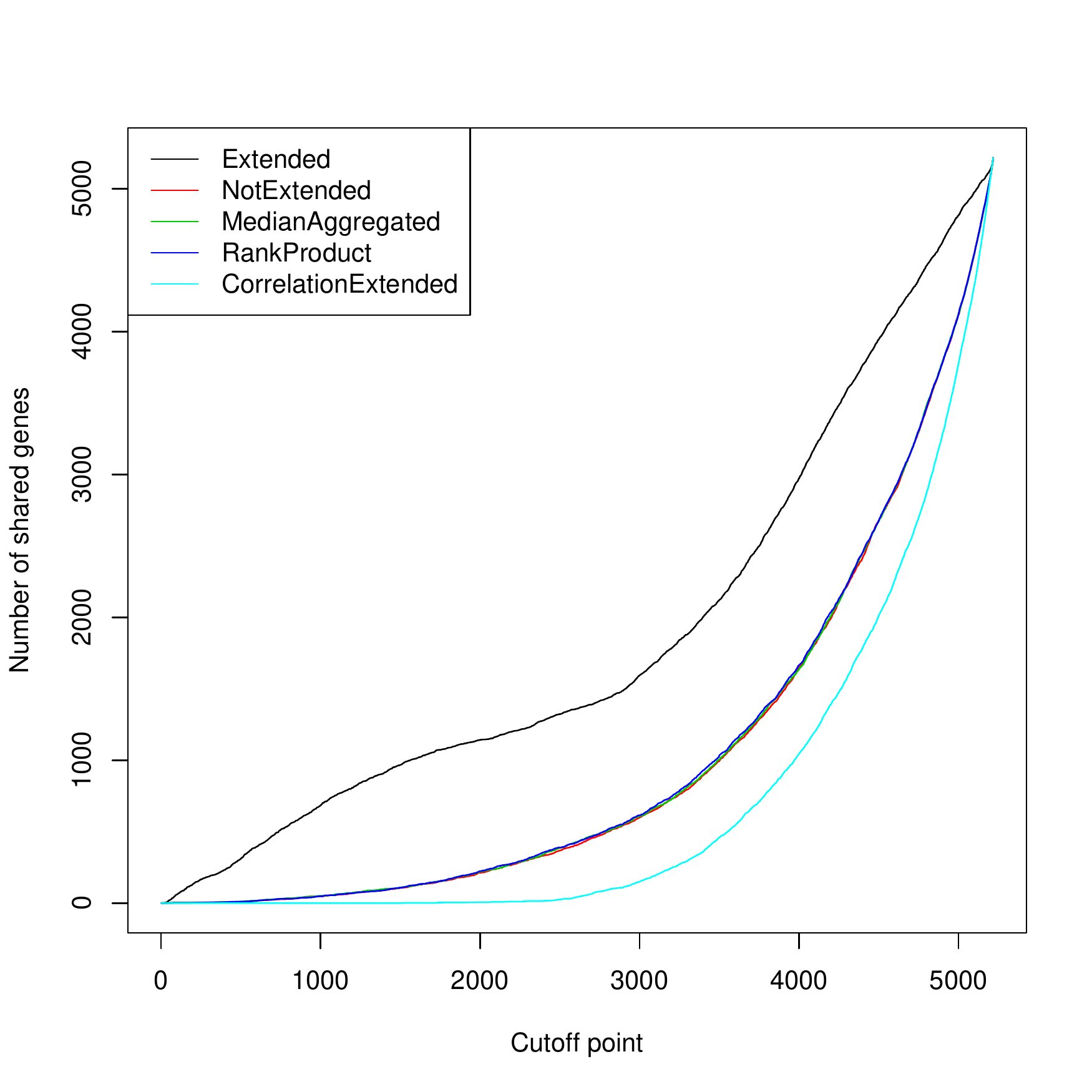}}}\\
\subfigure{\resizebox{55mm}{!}{\label{concordance:c}\includegraphics{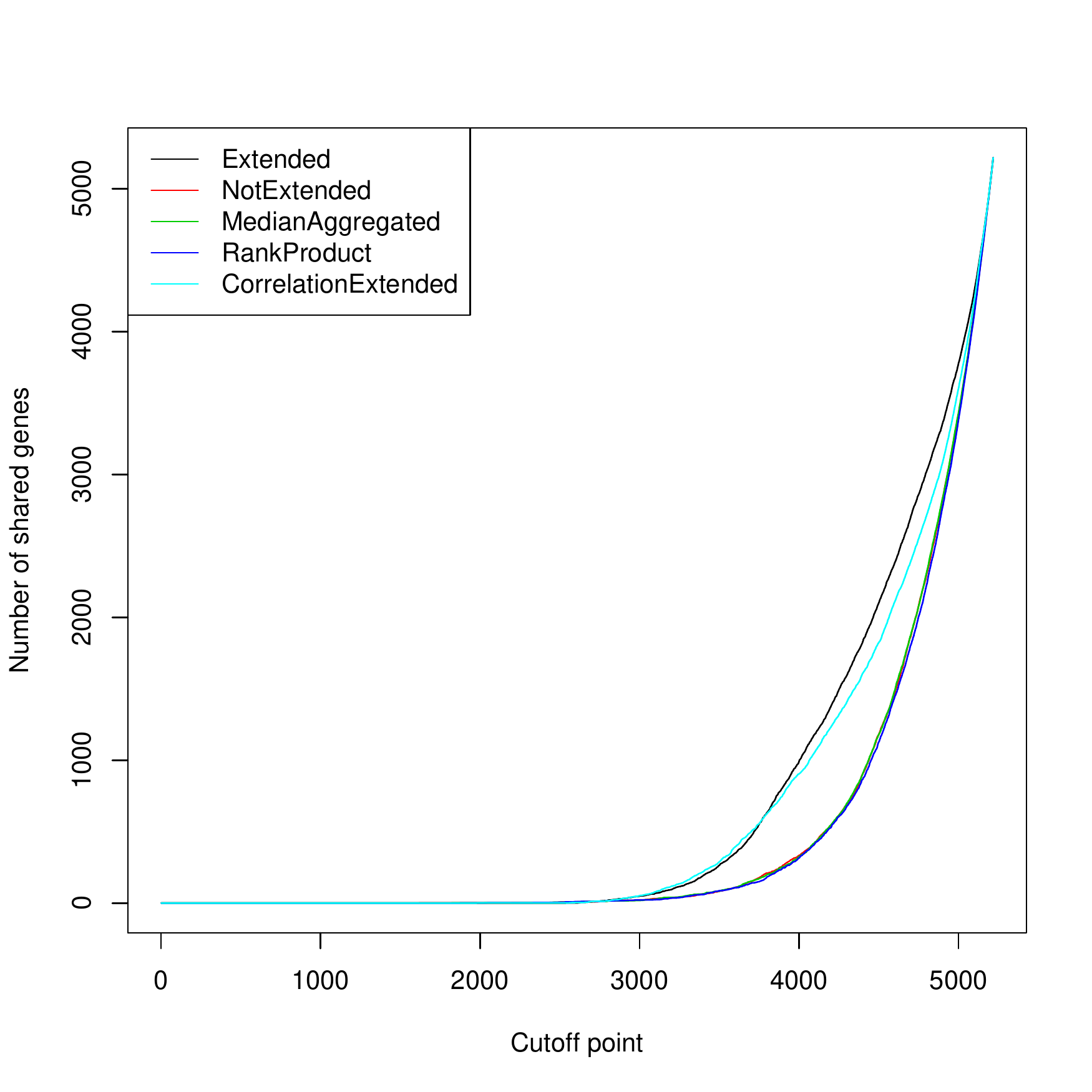}}}
\subfigure{\resizebox{55mm}{!}{\label{concordance:d}\includegraphics{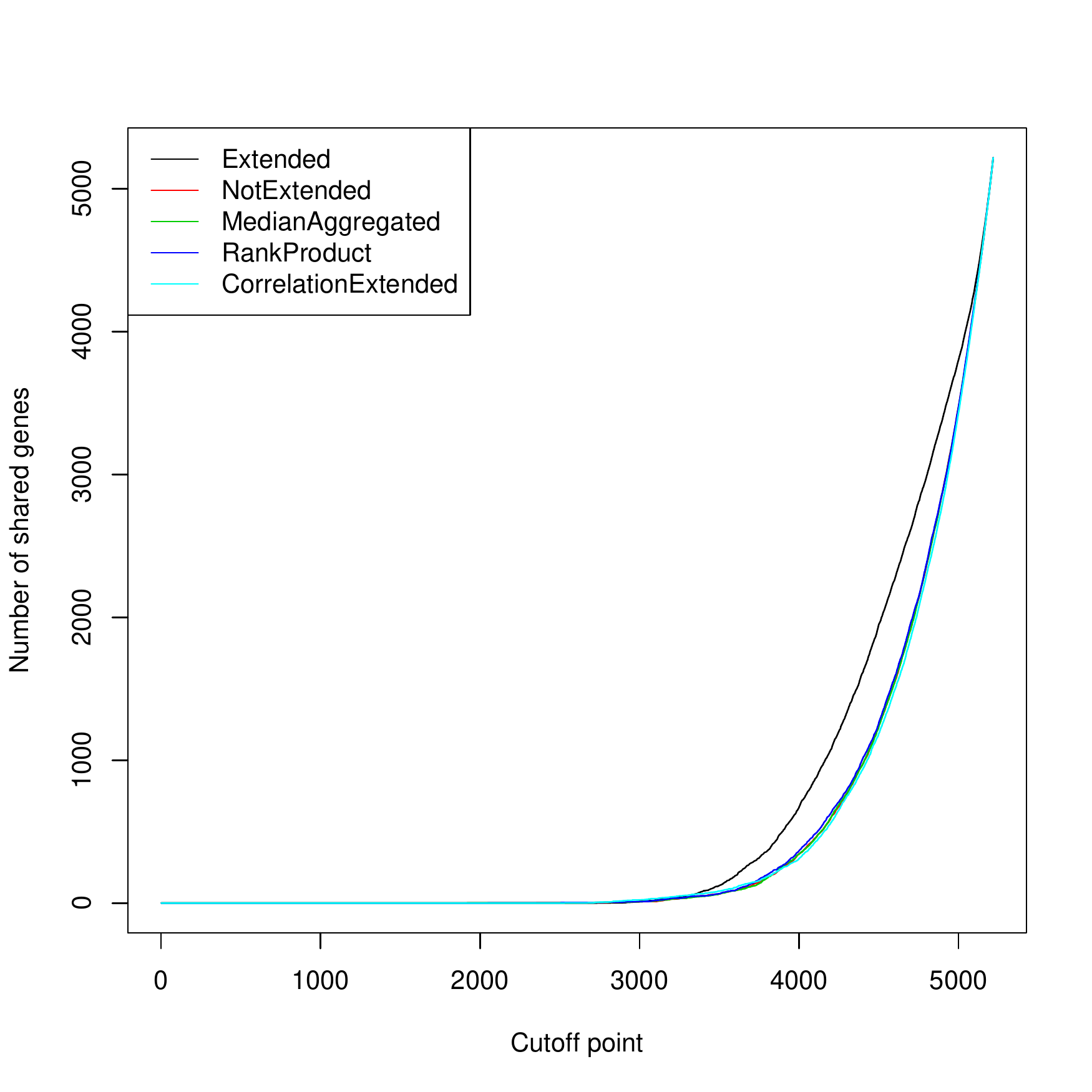}}}
\caption{Concordance plots for gene lists obtained by the five methods described in Section~\ref{sec:stability}. The top row shows concordance plots for the observed data (left panel: top genes, right panel: bottom genes) and the bottom row shows corresponding plots for data where the class labels have been randomly permuted, so that the gene rankings from different bootstrapped data set are unrelated. The curves corresponding to the original ranking and the two aggregated rankings coincide almost completely in all cases, indicating that the rankings obtained by aggregation are not more stable than the original rankings with respect to sampling variations in the underlying data set. The extended lists provide a more stable gene ranking for the observed data. Interestingly, using the positive part of the correlation matrix to extend the gene lists gives less stable rankings.}
\label{fig:concordance}
\end{center}
\end{figure}

\begin{figure}[hbt]
\centering
\resizebox{\textwidth}{!}{\includegraphics{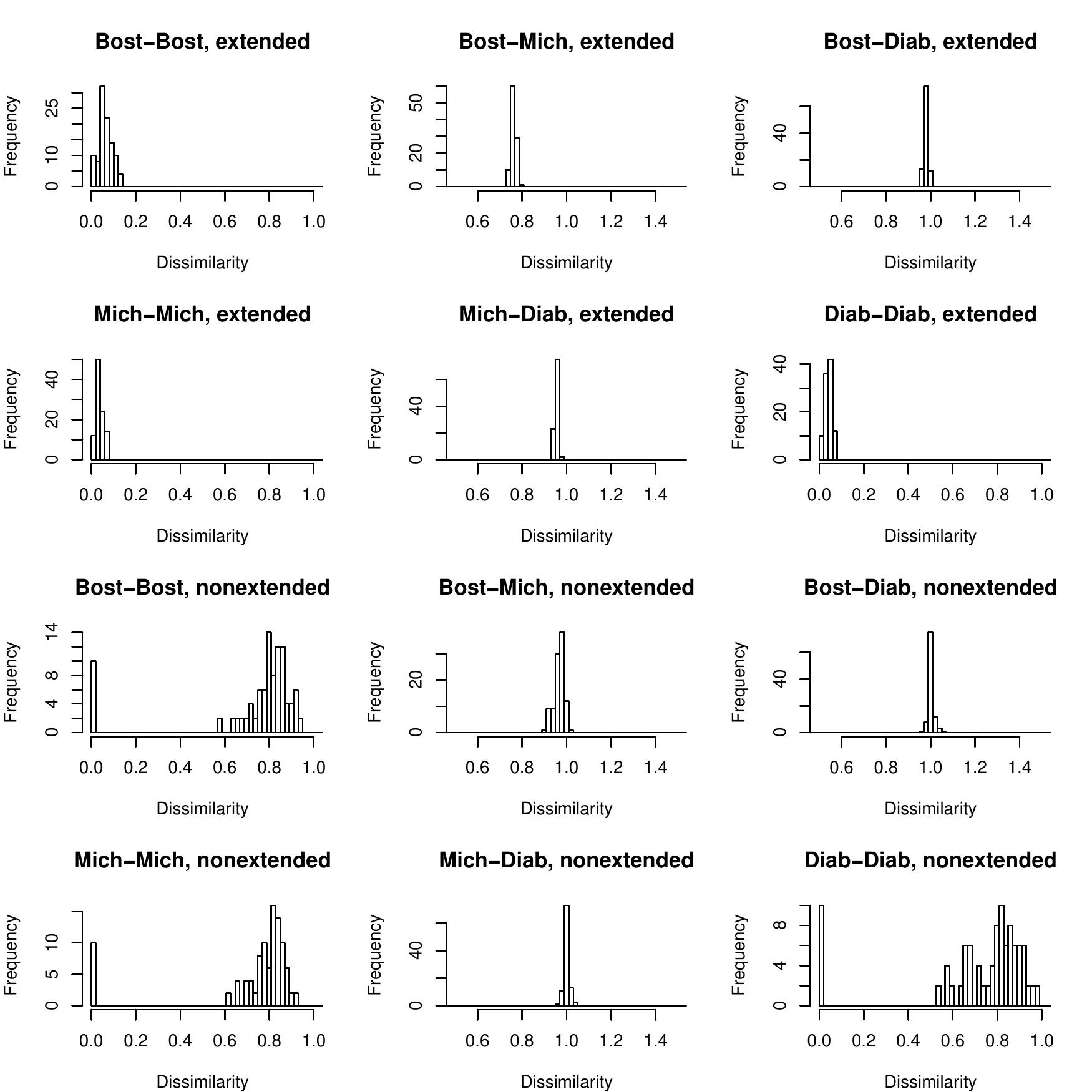}}
\caption{Histograms of the distances computed between ranked lists from the same data set or from different data sets with and without stabilization through exchangeability extension. Bost - Boston lung cancer data, Mich - Michigan lung cancer data, Diab - Diabetes data.}
\label{fig:distancestabilityrankedlists}
\end{figure}

\clearpage
\appendix

\section{Estimating exchangeability distances}
In the main article, we indicate how to estimate the exchangeability distances between genes based on subsampled data. This section provides a more thorough description. Recall that for any gene $g_i$ in the universal set $\mathcal{G}$, we define a random variable $X_i$ representing the position of the gene in the gene ranking from an experiment. We use subsampling to generate $B$ slightly modified data sets from which we extract gene rankings. The observed ranking positions for $g_i$, i.e. the samples of the random variable $X_i$, are collected in a position vector $S_i$. Note that the positions from the $B$ rankings must be placed in the same order in all position vectors. Combining two position vectors $S_i$ and $S_j$ provides samples of $X_i\times X_j$ and $X_j\times X_i$, which are used to estimate the exchangeability distance between $g_i$ and $g_j$. 

First, the {\em sample exchangeability plot} for the two genes is obtained by depicting in the same plot $S_{i\times j}=\left\{\left((S_i)_k,(S_j)_k\right)\right\}_{k=1}^B$ and $S_{j\times i}=\left\{\left((S_j)_k,(S_i)_k\right)\right\}_{k=1}^B$. If these point sets overlap completely, the two genes are considered to be weakly exchangeable. The estimates of $ED^{max}_{X_i\times X_j}$ and $ED^{mean}_{X_i\times X_j}$ are obtained by \begin{align*}\widehat{ED}^{max}_{X_i\times X_j}&=\frac{HD_\rho\left(\left\{((S_i)_k,(S_j)_k)\right\}_{k=1}^B,\left\{((S_j)_v,(S_i)_v)\right\}_{v=1}^B\right)}{\rho((1,1),(M,M))}\\\widehat{ED}^{mean}_{X_i\times X_j}&=\frac{\sum_{k=1}^B \frac{1}{B}\dist_\rho\left(\left\{((S_i)_k,(S_j)_k)\right\},\left\{((S_j)_v,(S_i)_v)\right\}_{v=1}^B\right)}{\rho((1,1),(M,M))}.\end{align*}We note that since $ED^{max}_{X_i\times X_j}$ only considers the largest distance between the two sets it is sensitive to outliers which may have detrimental effects when the number of samples used to estimate it ($B$) is small. To estimate the total exchangeability variation we first compute Gaussian kernel density estimates of the probability density functions of $X_i\times X_j$ and $X_j\times X_i$ from the samples $S_{i\times j}$ and $S_{j\times i}$. We then compute the difference between the two kernel density estimates and integrate the positive part (or, equivalently, the negative part) of the difference over $\mathbb{R}^2$. The value of the integral is taken as the estimate $\widehat{P}^{Var}_{X_i\times X_j}$. 
\begin{rem}Note that it is also possible to let $S_i$ be a binary vector such that $(S_i)_k=1$ if $g_i$ is present in some gene set output from the experiment (e.g. differentially expressed genes) and $(S_i)_k=0$ otherwise. \end{rem}

The one-sided exchangeability distance introduced in the main article penalizes gene pairs where one of the genes is always ranked before the other by computing distances only between points on the same side of the line $\{(x,y)\in\mathbb{R}^2;\,y=x\}$. 
Given two position vectors $S_i$ and $S_j$ of length $B$, we define $$R_{ij}(k)=\{(x,y)\in\mathbb{M}\times\mathbb{M};\,\,\sign\left(x-y\right)=\sign\left(\left(S_i\right)_k-\left(S_j\right)_k\right)\}$$for $k=1,\ldots,B$. The one-sided mean exchangeability distance between the two genes is then estimated by $$\widehat{oED}^{mean}_{X_i\times X_j}=\frac{\sum_{k=1}^B\frac{1}{B}\dist_\rho\left(\left\{\left((S_i)_k,(S_j)_k\right)\right\},\left\{((S_j)_v,(S_i)_v)\right\}_{v=1}^B\bigcap R_{ij}(k)\right)}{\rho\left(\left(1,2\right),\left(M-1,M\right)\right)}$$ if $\left\{\left((S_j)_v,(S_i)_v\right)\right\}_{v=1}^B\cap R_{ij}(k)\neq\emptyset$ for all $k$, 
and we define $\widehat{oED}^{mean}_{X_i\times X_j}=1$ otherwise. 

To estimate normalized exchangeability scores, e.g. $$\widehat{noES}^{mean}_{X_i\times X_j}=\left(\frac{\widehat{oES}^{mean}_{X_i\times X_j}-\widehat{oES}^{mean}_{Y_i\times Y_j}}{1-\widehat{oES}^{mean}_{Y_i\times Y_j}}\right)_+,$$for the Boston lung cancer data we note that for virtually all pairs of variables $(X_i,X_j)$, the cardinality of the observed support of $X_i\times X_j$ is equal to $B$. Therefore, to compute $\widehat{oES}^{mean}_{Y_i\times Y_j}$, we repeatedly sample $B$ points uniformly from $\mathbb{M}\times\mathbb{M}$ to form $\supp Y_i\times Y_j$, and compute the resulting exchangeability scores for $Y_i\times Y_j$ uniformly distributed on its support. Then $\widehat{oES}^{mean}_{Y_i\times Y_j}$ is taken to be the mean value of the estimated exchangeability scores. 

\section{Constructing the position matrix $A_\ell$}
In the main article we use the rank-based function $$(A_\ell)_{ii}=\left\{\begin{array}{cl}\frac{b^2}{(\pi_\ell(i)-1)^2+b^2}&\textrm{if }SNR(i)\geq 0\\-\frac{b^2}{(M-\pi_\ell(i))^2+b^2}&\textrm{if }SNR(i)<0,\end{array}\right.$$with $b^2=350$ and $M=5,217$ to construct the position matrix $A_\ell$ from the positions of the genes in the ranking. This function and parameters were chosen heuristically, and here we give some motivation behind the choices. First, we want to treat the variables which are positively and negatively associated to poor outcome symmetrically, hence the different expressions for positive and negative SNR. Second, we want the position values to be somewhat adapted to the exchangeability scores. If the position values decrease too fast, the influence of the exchangeabilities is too high since all variables exchangeable with the top gene will be moved close to the top of the list. On the other hand, if the position values decrease too slowly, the exchangeabilities only have a marginal impact. We found the selected function to provide a good trade-off between these two effects.

We chose a rank-based function since this type of functions can be used also when there is no clearly defined ranking statistic. Note that another way of constructing the position matrix would have been to choose $(A_\ell)_{ii}=SNR(i)$, which in this case gives very similar results (data not shown). 

\section{Choosing the weight matrix $W_\ell$}
The global weight matrix $W_\ell$ in the general framework presented in the main article can be used to incorporate any information about the quality of the measurements corresponding to each gene. In the examples in the paper, we do not have this information and hence use $W_\ell=I_M$. Here, we give some examples of how the weight matrix can be chosen. 
Given a collection of $K$ reference lists, we can choose e.g. $$(W_\ell)_{ii}=\log\left(\frac{K+1}{K_i+1}\right)$$ where $K_i$ is the number of reference lists containing $g_i$. This resembles the inverse document frequency weighting commonly used in the field of information retrieval. With this choice, a gene which is present in a large number of lists obtains a lower weight. Another weighting scheme could take into account the stability of the position of the gene in the list, which could be estimated through resampling techniques. A gene which is often in the same position would then have a large influence on the list vector relative to a gene which can be in many different positions.

\section{Relationship between positive part of correlations and exchangeabilities in the Boston lung cancer data}
\begin{figure}[hbt]
\centering
\resizebox{100mm}{!}{\includegraphics{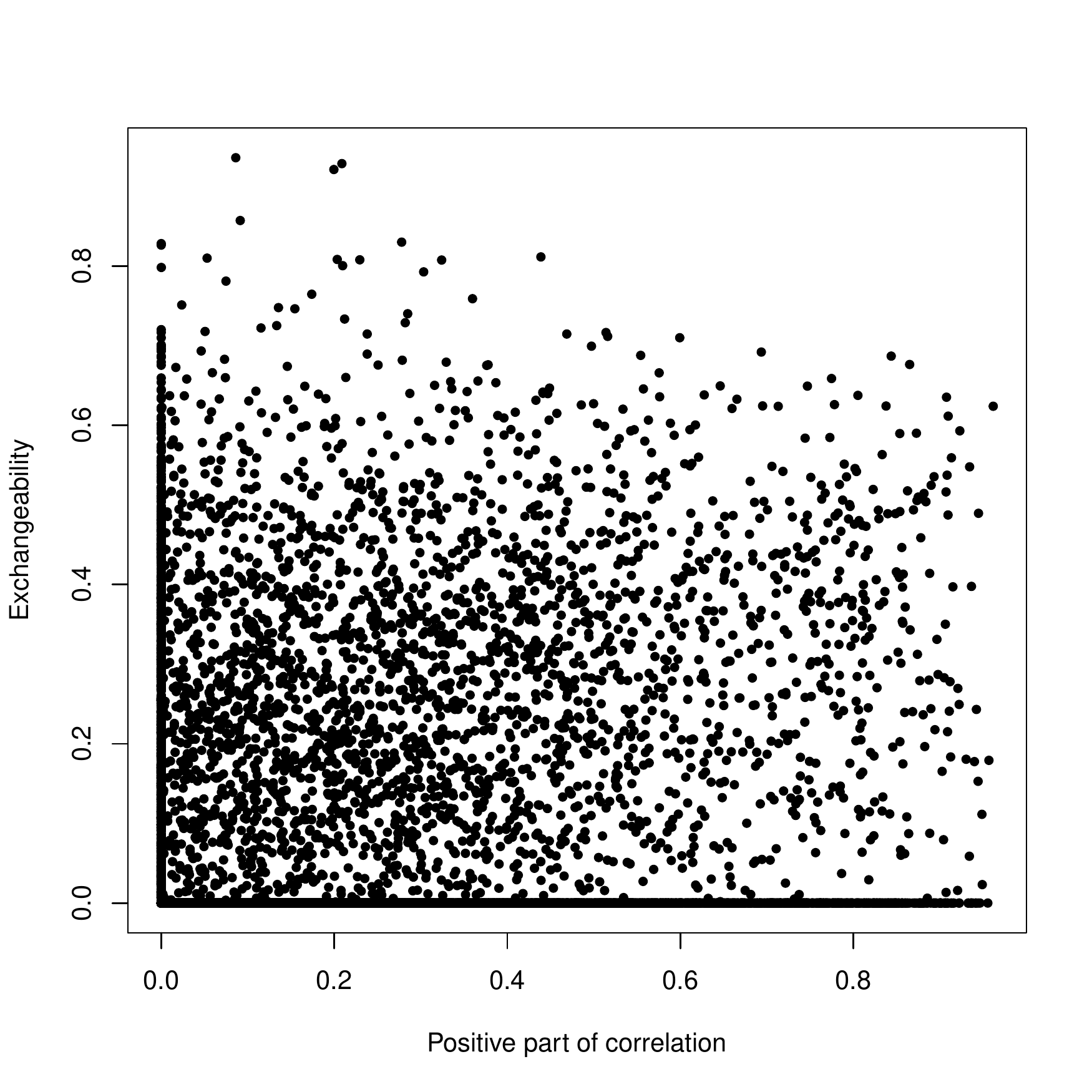}}
\caption{Positive part of correlation coefficient between expression values and the estimated exchangeabilities for 10,000 randomly chosen gene pairs from the Boston lung cancer data. }
\label{fig:correxchboston}
\end{figure}

As discussed in the main article, using the positive part of the correlation matrix to extend the list vector did not give more stable rankings. Figure~\ref{fig:correxchboston} shows the relationship between the positive part of the correlation between the expression values and the estimated exchangeabilities for 10,000 randomly chosen gene pairs in the Boston lung cancer data. Apparently, in this data set the correlation between expression values has little to do with the estimated exchangeability between the genes. 

\section{Overlap plots}\label{app:overlap30}
\begin{figure}[htb]
\begin{center}
\subfigure{\resizebox{100mm}{!}{\label{overlaps:a}\includegraphics{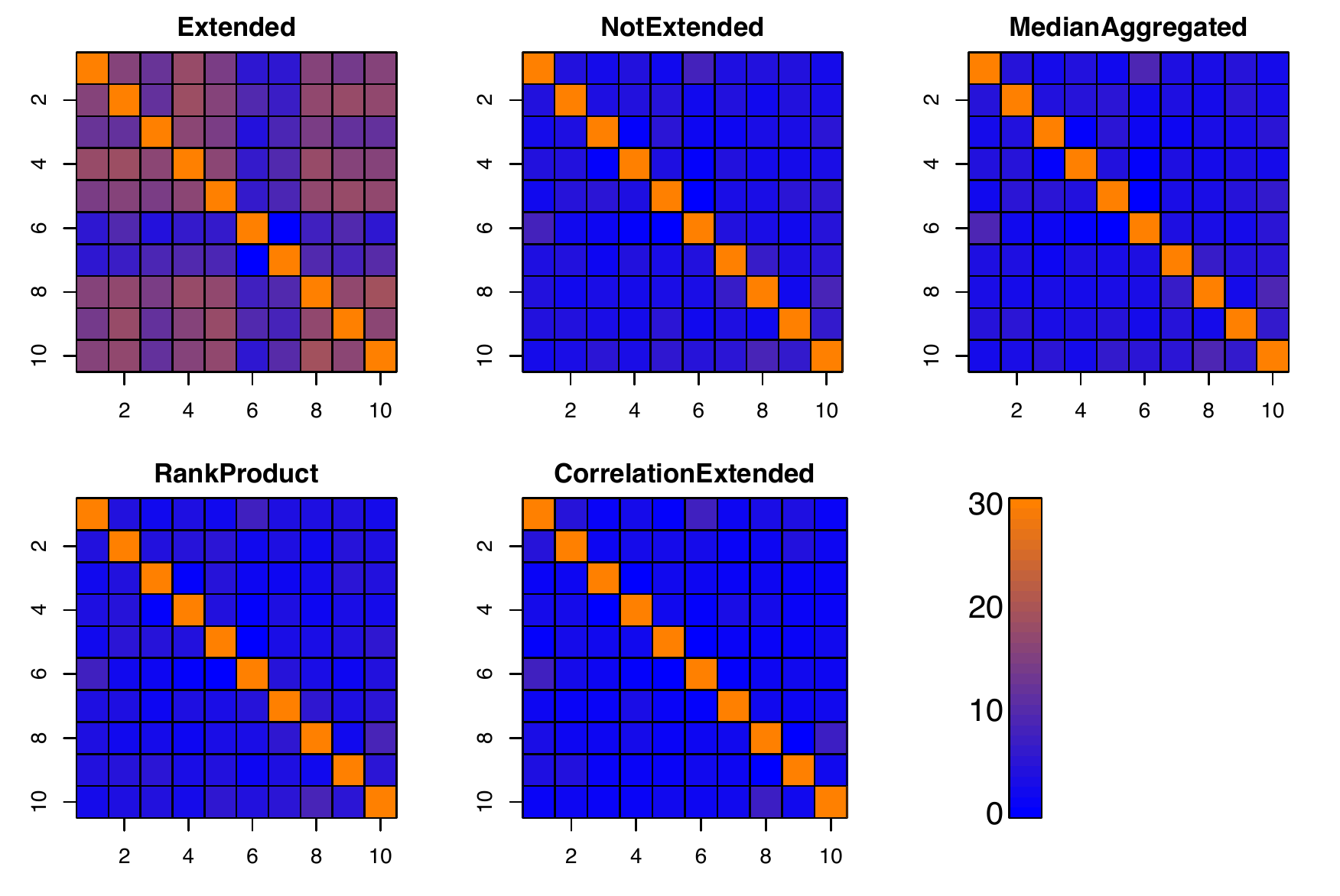}}}
\subfigure{\resizebox{100mm}{!}{\label{overlaps:b}\includegraphics{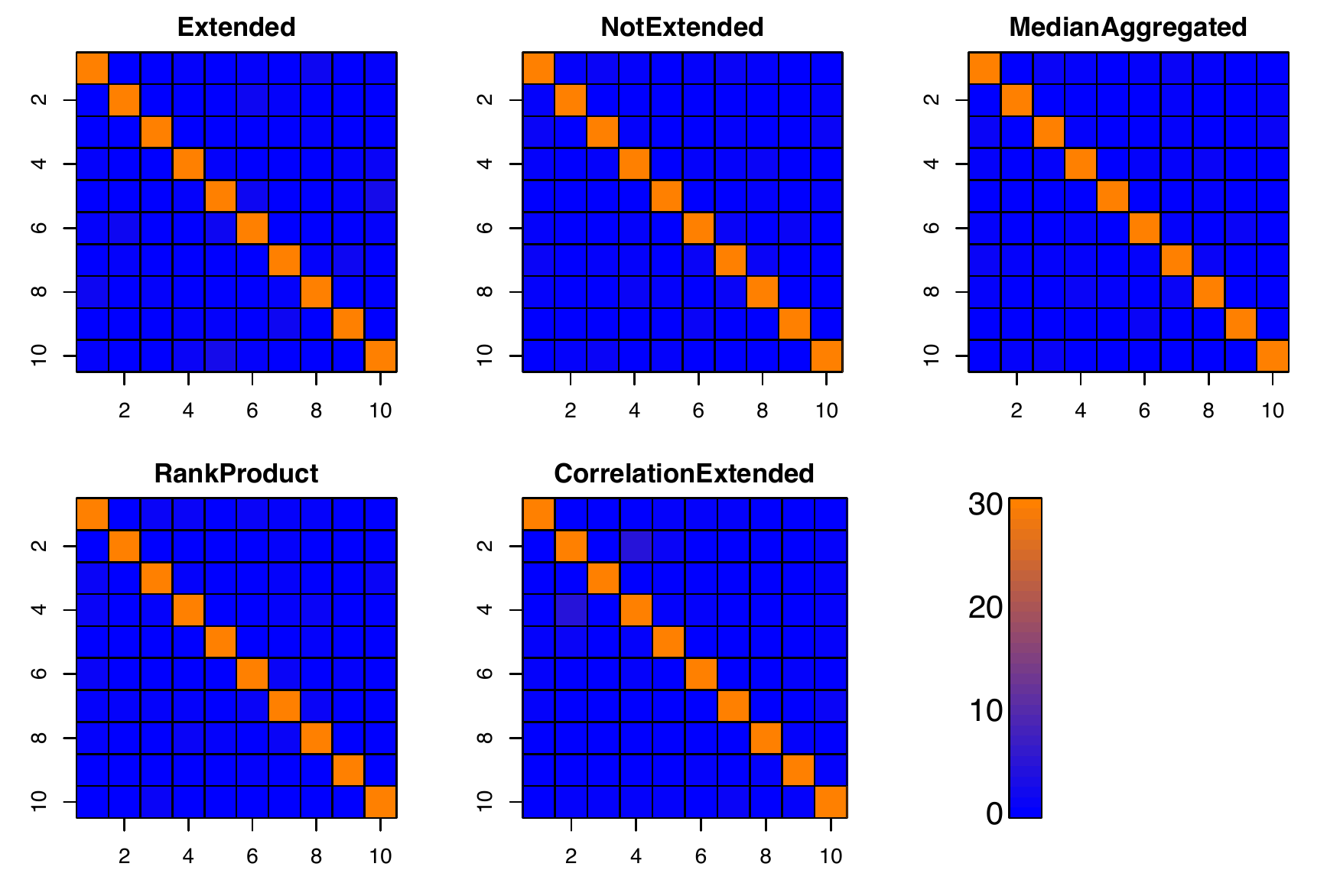}}}
\caption{Top rows: The mean overlap between the top-30 and bottom-30 genes in the lists obtained by the five ranking methods described in the main article, for each pair of bootstrapped Boston lung data sets. The overlap between the top and bottom genes is larger for extended lists than for lists obtained by the other methods. Bottom rows: Corresponding plots for lists extracted from data sets with permuted class labels. In this case there is no functional connection between the lists which is captured by the almost non-existent overlap between the top genes in all comparisons.}
\label{fig:overlapbostonlungbootstrap}
\end{center}
\end{figure}

The concordance plots shown in the main article give a visualization of the overlap among the rankings from {\em all} the 10 bootstrapped data sets. Here, we show overlap plots, giving the mean overlap between the top- and bottom-30 genes from each pair of bootstrapped data sets, for each of the ranking metrics. This emphasizes other aspects of the stability since we only consider pairwise overlaps. 
The top rows in Figure~\ref{fig:overlapbostonlungbootstrap} show the mean overlap between the top-30 and bottom-30 genes for each pair of bootstrapped data sets. It is clear that the top parts of the extended lists have a larger overlap than the non-extended or aggregated lists, further indicating that the exchangeability captures relevant biological information. The bottom rows in Figure~\ref{fig:overlapbostonlungbootstrap} show corresponding plots for random rankings (obtained by independently permuting the sample labels in each bootstrapped data set). These figures show that in the absence of a functional connection between two gene lists, no lists have the same top-ranked genes.

\section{Distance between ranked lists from permuted data}
\begin{figure}[hbt]
\centering
\resizebox{\textwidth}{!}{\includegraphics{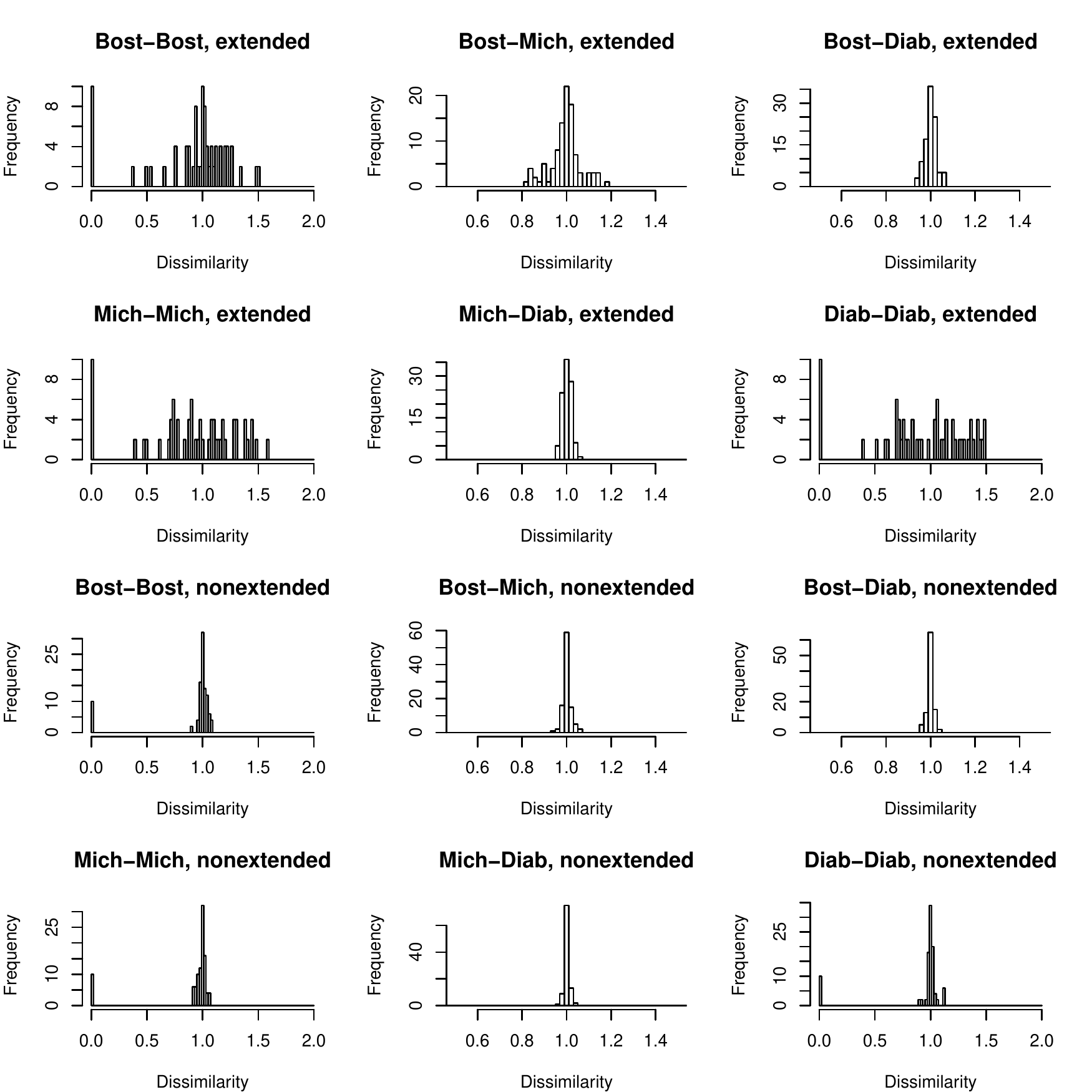}}
\caption{Histograms of the distances computed between ranked lists from the same data set or from different data sets with and without stabilization through exchangeability extension, for random rankings. Bost - Boston lung cancer data, Mich - Michigan lung cancer data, Diab - Diabetes data.}
\label{fig:distancestabilityrankedlistsrandom}
\end{figure}

In the main article, we studied the stability of distances between ranked lists from different experiments (Figure~3 in the main article). Figure~\ref{fig:distancestabilityrankedlistsrandom} shows corresponding plots for random rankings, i.e. for rankings obtained from data sets were the class labels are permuted independently in each bootstrap round. The distance between the extended list vectors are not more stable than the distance between the non-extended list vectors and as expected, the list vectors from the same data set are more dissimilar than for the original data (Figure~3 in the main article), since there is no shared information between the different rankings. 

\section{Interpretation of similarity score}\label{sec:simscoreinterpretation}
In this section we show how to find the genes which are the most important for explaining the similarity between two gene lists. As an example, we take the lists obtained from the Boston lung cancer data and the Michigan lung cancer data. We compute extended lists as described in the main article, using $\widehat{noES}^{mean}_{X_i\times X_j}$ as a measure of exchangeability in each data set. Then we compute the dot product between the resulting extended list vectors and find the genes with the highest contribution. Table~\ref{tab:genecontribution} shows the genes with the highest contribution, their position in the original (non-extended) list and their position in the extended lists from each data set. 

\begin{table}[htb]
\caption{Genes with highest influence on the similarity between the extended list vectors from Boston lung data and Michigan lung data.}
\centering
\resizebox{0.95\textwidth}{!}{
\begin{tabular}{llllll}
&&&&&\\
&&Position in&Position in&Position in&Position in\\
Gene&Contribution&ext. list Boston&ext. list Michigan&nonext. list Boston&nonext. list Michigan\\
\hline
CASP4&0.922&5,207&5,208&5,198&5,191\\
DBP&0.910&1&19&1&40\\
ENO2&0.837&5,217&5,165&5,217&5,186\\
FADD&0.836&5,162&5,215&5,141&5,215\\
KRT18&0.826&5,196&5,183&5,205&5,206\\
CR2&0.826&29&8&13&17\\
TUBA1&0.819&5,194&5,184&5,199&5,185\\
LAMB3&0.798&5,168&5,195&5,168&5,164\\
KRT19&0.794&5,160&5,206&5,140&5,204\\
BZW1&0.780&5,201&5,158&5,209&5,181
\end{tabular}
}
\label{tab:genecontribution}
\end{table}

\section{Synthetic examples}\label{app:examples}
In this section we give some examples comparing the different measures of exchangeability to each other and to the correlation coefficient, which is often used to quantify the relationship between two genes. We expect the exchangeability to highlight other relationships than the correlation coefficient since the latter does not take into account the specific experiment. 

\begin{ex}\label{ex:overexpressed}
We simulate a synthetic data matrix $X\in\mathbb{R}^{50\times 40}$ by letting $$X_{ij}\in\left\{\begin{array}{ll}\mathcal{N}(1,1)&\textrm{if }\,1\leq i\leq 10, 1\leq j\leq 20\\\mathcal{N}(0,1)&\textrm{otherwise.}\end{array}\right.$$As the variable ranking method we use a univariate t-test contrasting the first 20 samples against the last 20 samples. Clearly, with respect to this experiment the first 10 variables should be highly exchangeable since on a population level, they are all equally related to the contrast between the two sample groups. Similarly, there should also be a certain degree of exchangeability between the last 40 variables, since none of them is related to the response. We generate position vectors for the genes by subsampling the original data set $B=50$ times, each time keeping 2/3 of the samples from each group. 
We then compute $\widehat{nPS}^{Var}_{X_i\times X_j}$, $\widehat{nES}^{mean}_{X_i\times X_j}$, $\widehat{nES}^{max}_{X_i\times X_j}$, $\widehat{noES}^{mean}_{X_i\times X_j}$, $\widehat{noES}^{max}_{X_i\times X_j}$ and the positive part of the correlation coefficient between each pair of variables. 
Figure~\ref{fig:significantexchangeabilitiesoverexpressed} shows the exchangeability matrices and the positive part of the correlation matrix, averaged over 10 realizations. All exchangeability measures clearly divide the genes into two groups consisting of the first 10 and the last 40 genes, with only very few nonzero exchangeabilities between genes from different groups. Moreover, the first 10 genes are more highly exchangeable than the last 40. The correlations do not detect this structure.  

\begin{figure}[!htb]
\begin{center}
\subfigure[]{\resizebox{55mm}{55mm}{\label{fig:overexpressedexch1}\includegraphics{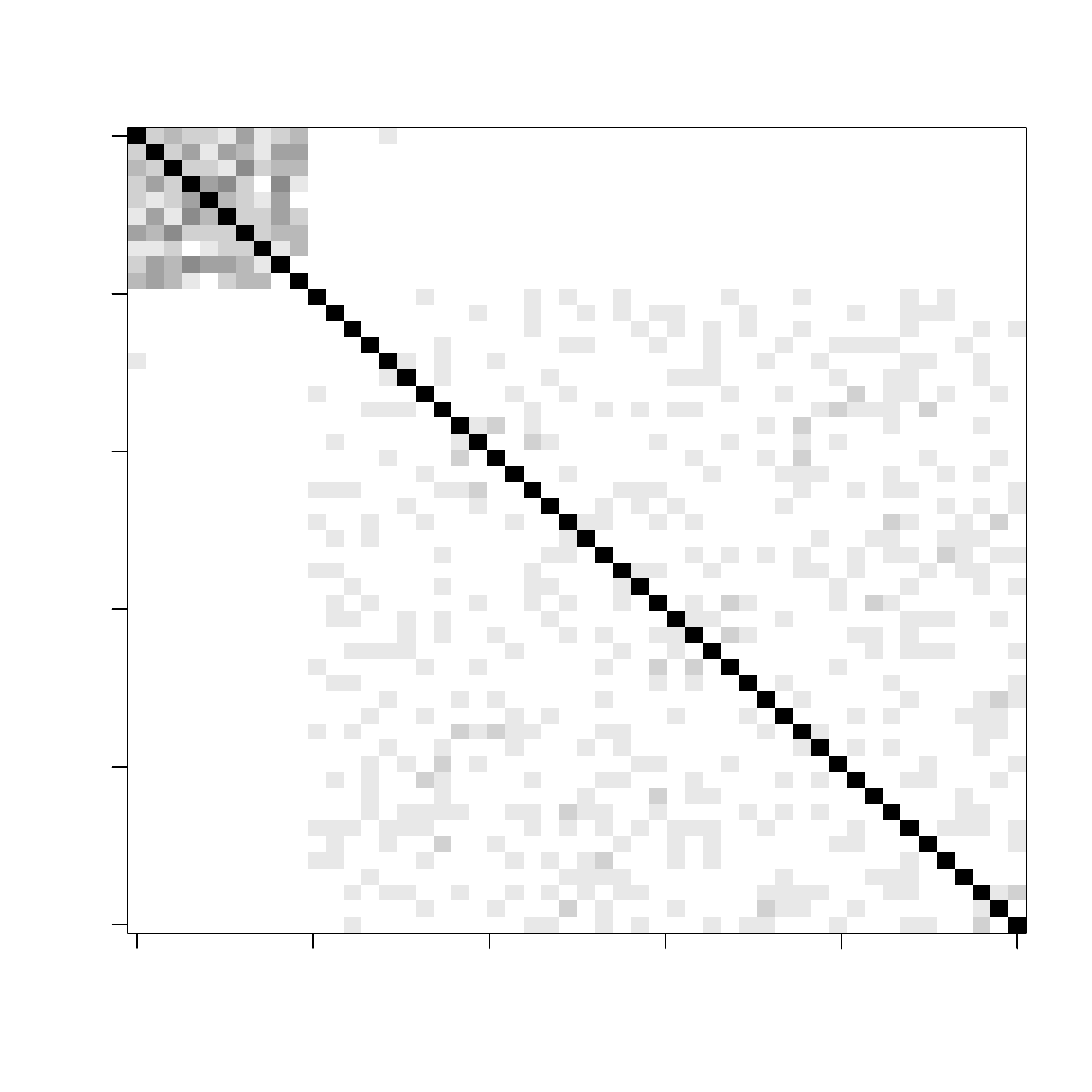}}}
\subfigure[]{\resizebox{55mm}{55mm}{\label{fig:overexpressedexch2}\includegraphics{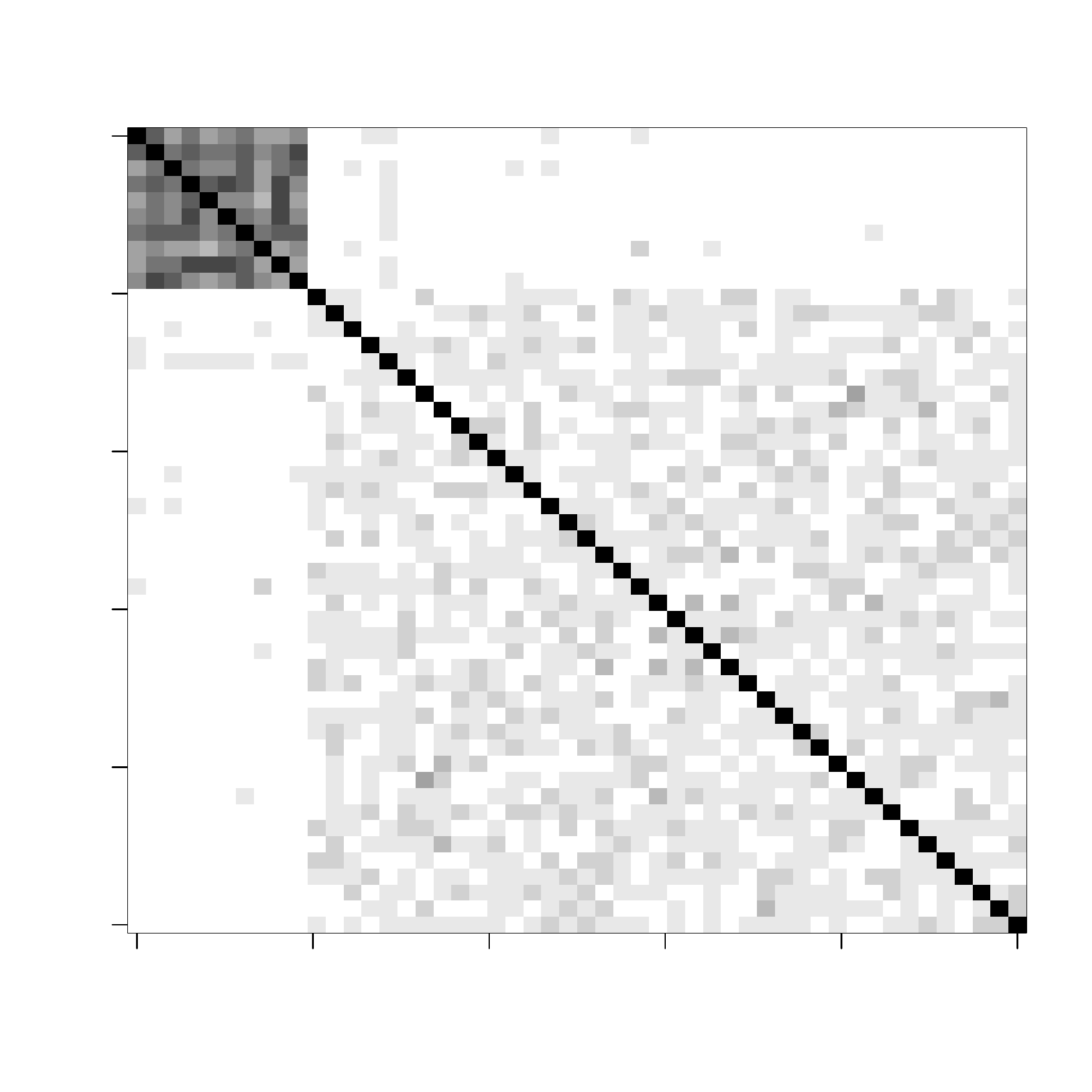}}}
\subfigure[]{\resizebox{55mm}{55mm}{\label{fig:overexpressedexch3}\includegraphics{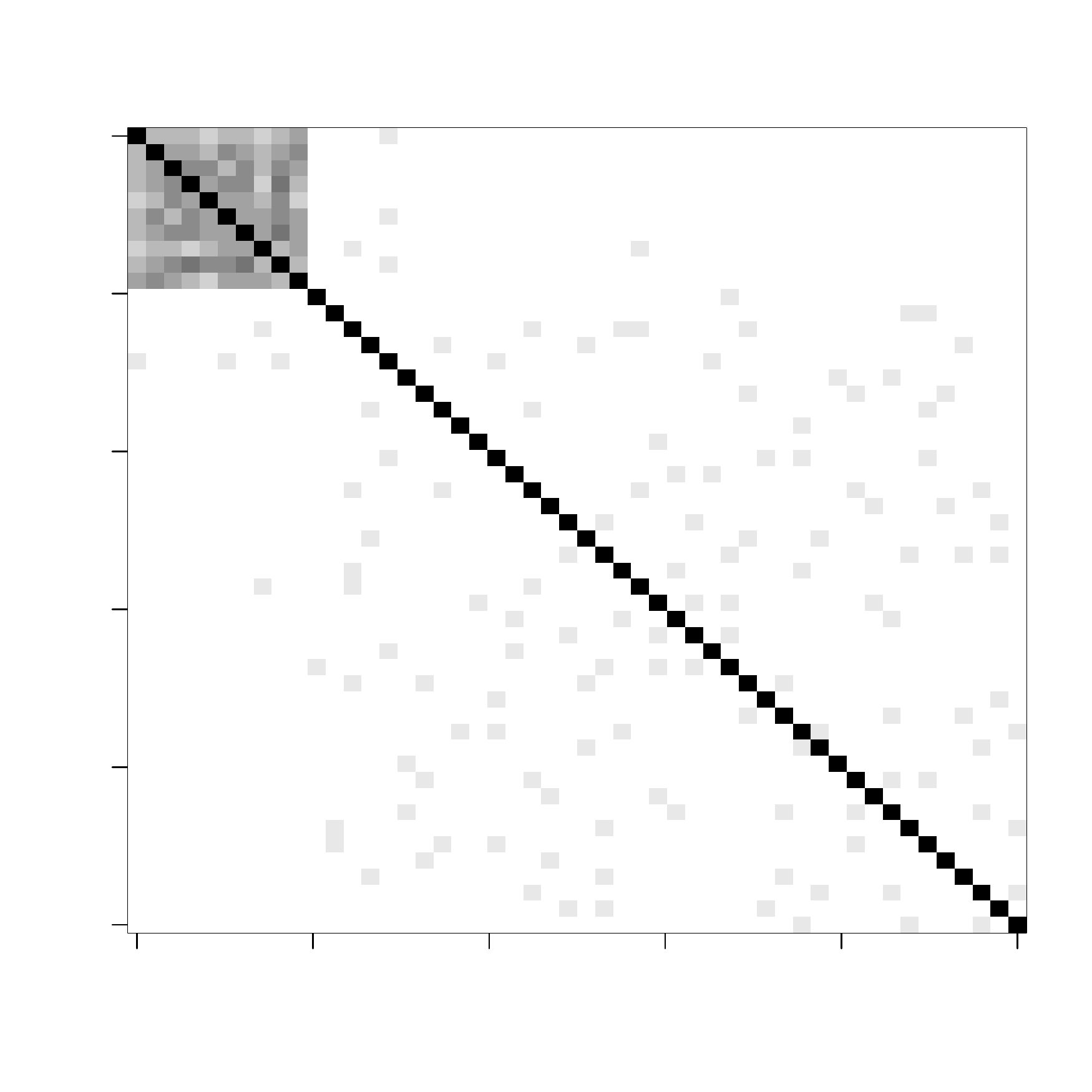}}}
\subfigure[]{\resizebox{55mm}{55mm}{\label{fig:overexpressedexch4}\includegraphics{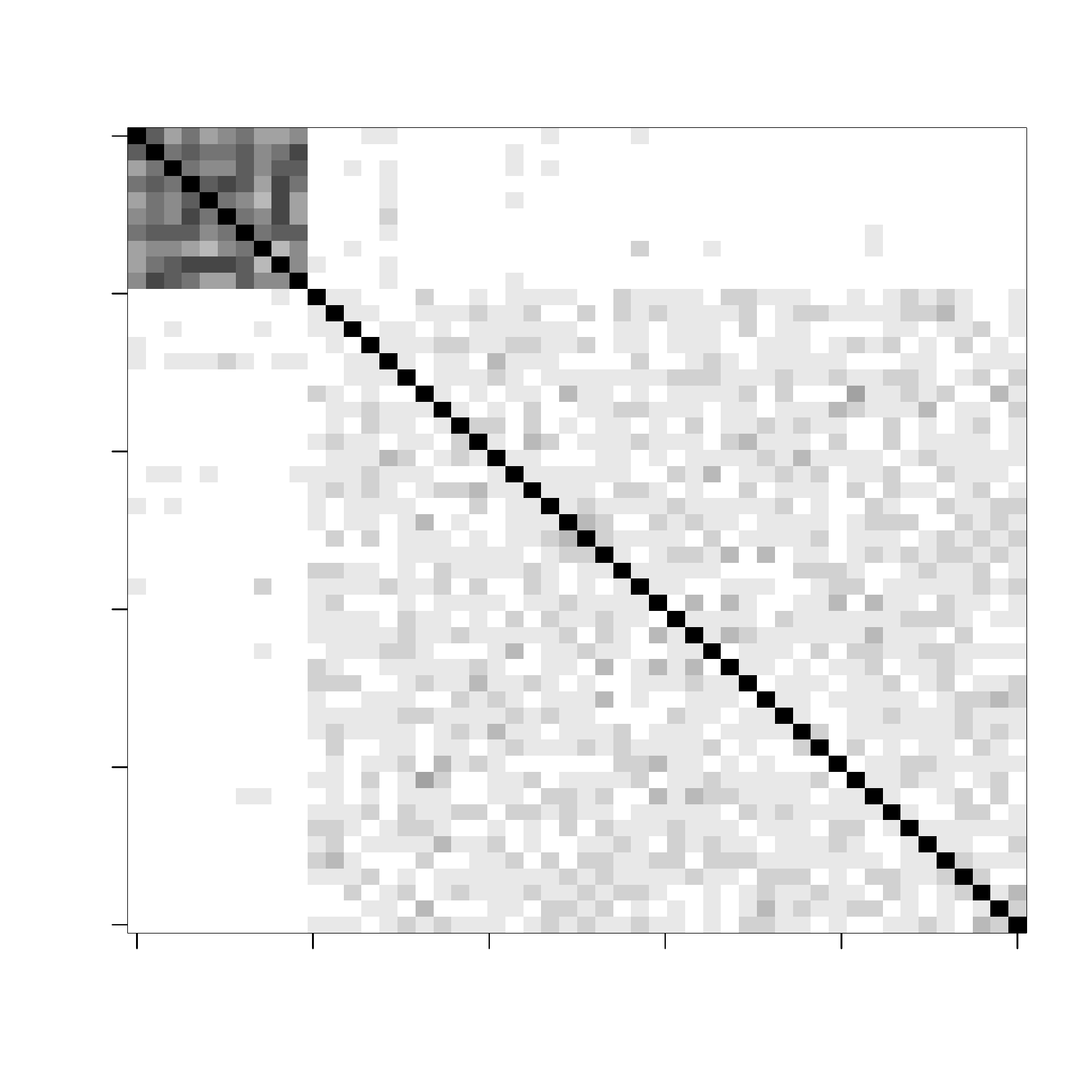}}}
\subfigure[]{\resizebox{55mm}{55mm}{\label{fig:overexpressedexch5}\includegraphics{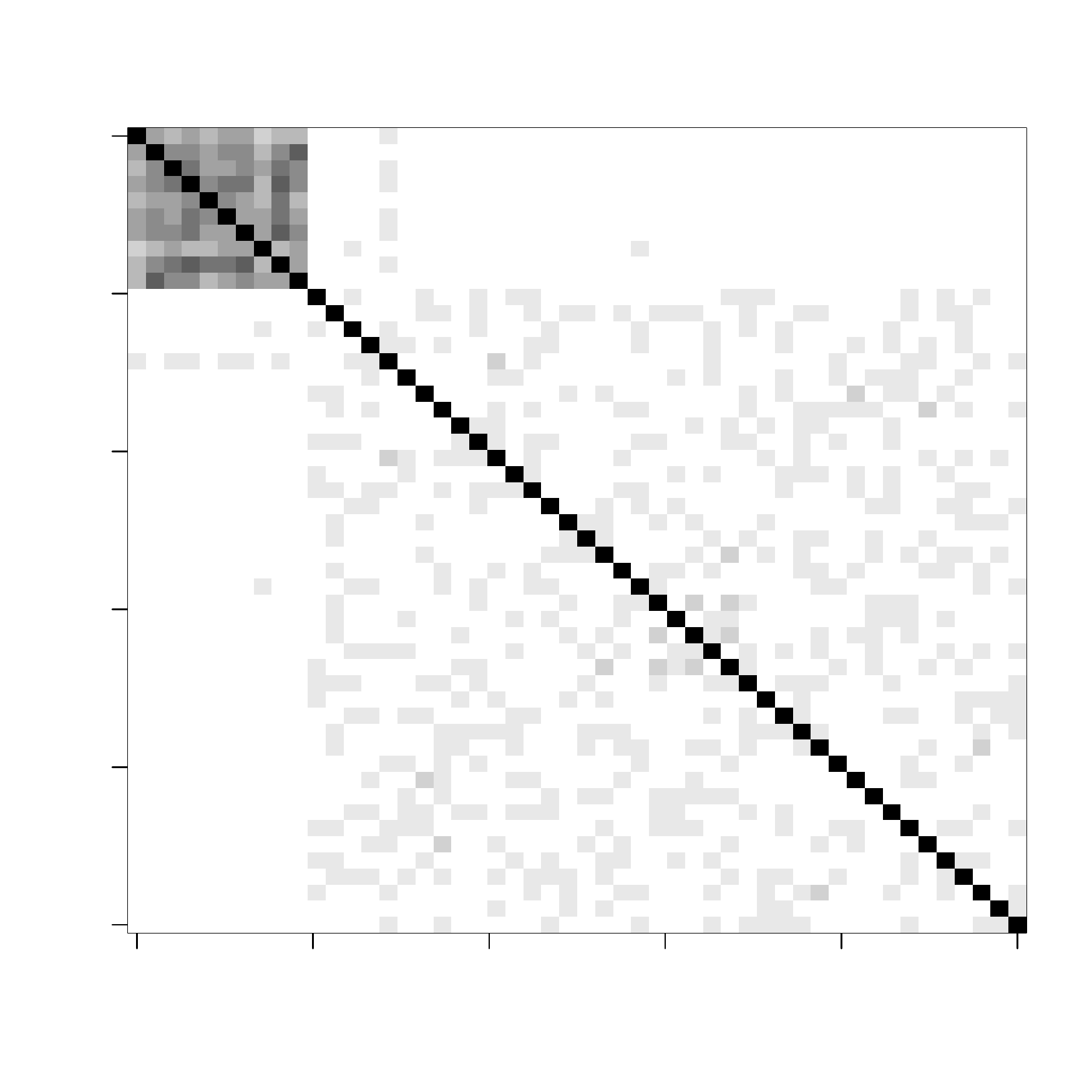}}}
\subfigure[]{\resizebox{55mm}{55mm}{\label{fig:overexpressedcorr}\includegraphics{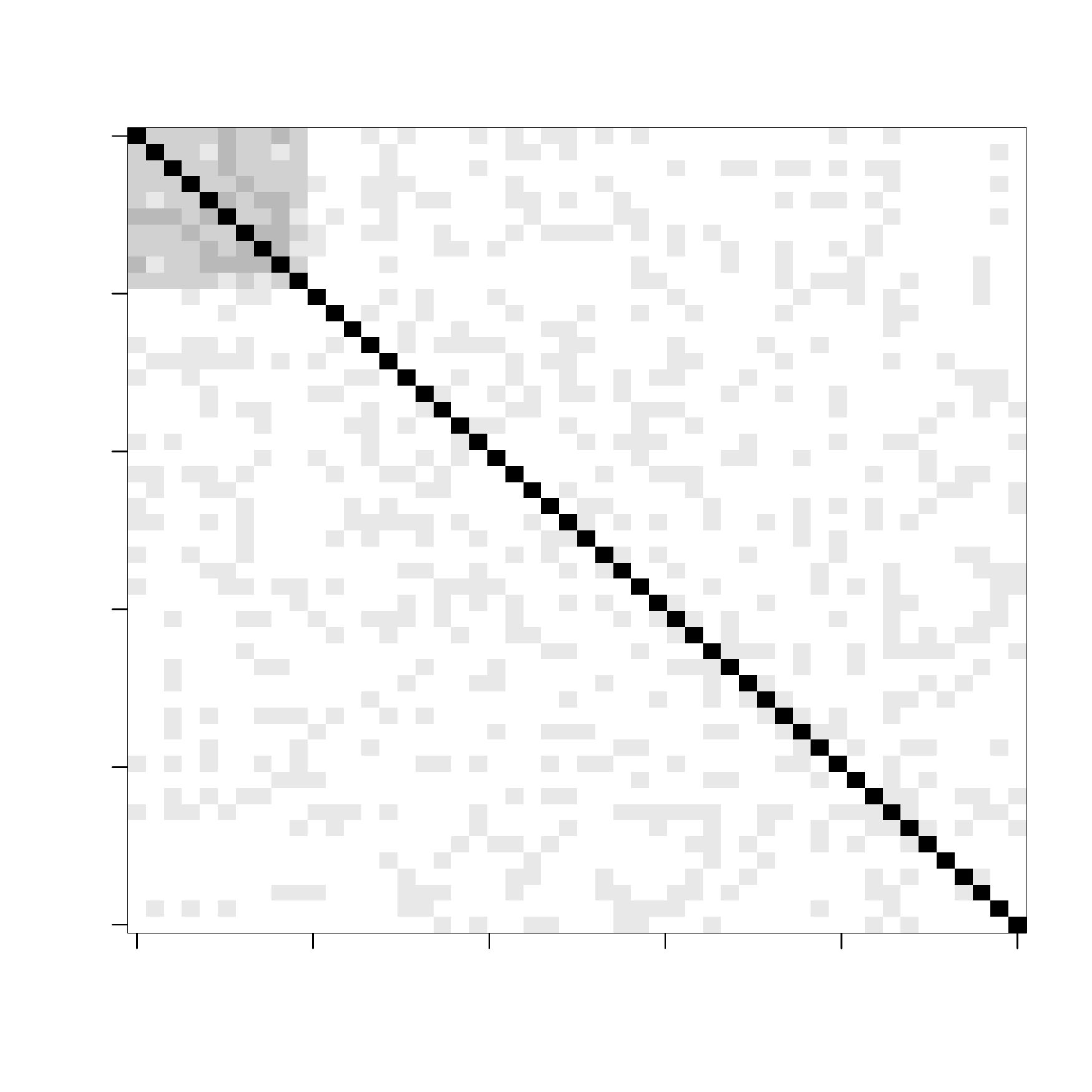}}}
\caption{Exchangeabilities and positive part of correlation for variables in Example~\ref{ex:overexpressed}, averaged across 10 simulations. 
(a) $\widehat{nPS}^{Var}_{X_i\times X_j}$. 
(b) $\widehat{nES}^{mean}_{X_i\times X_j}$. 
(c) $\widehat{nES}^{max}_{X_i\times X_j}$. 
(d) $\widehat{noES}^{mean}_{X_i\times X_j}$.
(e) $\widehat{noES}^{max}_{X_i\times X_j}$.
(f) positive part of correlations.  }
\label{fig:significantexchangeabilitiesoverexpressed}
\end{center}
\end{figure}
\end{ex}

\begin{ex}\label{ex:coregulated}We simulate a synthetic data matrix $X\in\mathbb{R}^{75\times 60}$ such that $$X_{ij}\in\left\{\begin{array}{ll}\mathcal{N}(2,1)&\textrm{if }\,1\leq i\leq 8, 1\leq j\leq 15\\\mathcal{N}(2,1)&\textrm{if }\,9\leq i\leq 16, 16\leq j\leq 30\\\mathcal{N}(0,1)&\textrm{otherwise.}\end{array}\right.$$We assign the first 30 samples to one group and the next 30 samples to another group. Hence, the first group of 8 variables and the next group of 8 variables are both related to the contrast between the two groups, but the two groups of variables are mutually exclusively overexpressed in each sample. This could correspond to a situation where these two groups of variables have the same function in the cell and therefore do not all have to be overexpressed in a particular sample. The structure of the data matrix is shown in Figure~\ref{fig:coregulateddata}. We rank the variables by comparing the two sample groups with a univariate t-test. We estimate the exchangeability scores $\widehat{noES}^{mean}_{X_i\times X_j}$ between the variables from position vectors obtained by subsampling the data set $B=50$ times, each time keeping 2/3 of the samples from each group. 
Figure~\ref{fig:coregulatedexchangeability} shows the exchangeability matrix and Figure~\ref{fig:coregulatedcorrelation} shows the positive part of the correlations, all averaged over 10 realizations. The exchangeability score detects the equivalence of the two groups of 8 variables with respect to the ranking method. The correlation does not take the response into account and hence does not find the relationship between the two groups of variables. Figure~\ref{fig:coregulatedcorrexch} shows the exchangeability scores of the variable pairs plotted against the corresponding correlations.

\begin{figure}[!htb]
\begin{center}
\subfigure[]{\resizebox{55mm}{55mm}{\label{fig:coregulateddata}\includegraphics{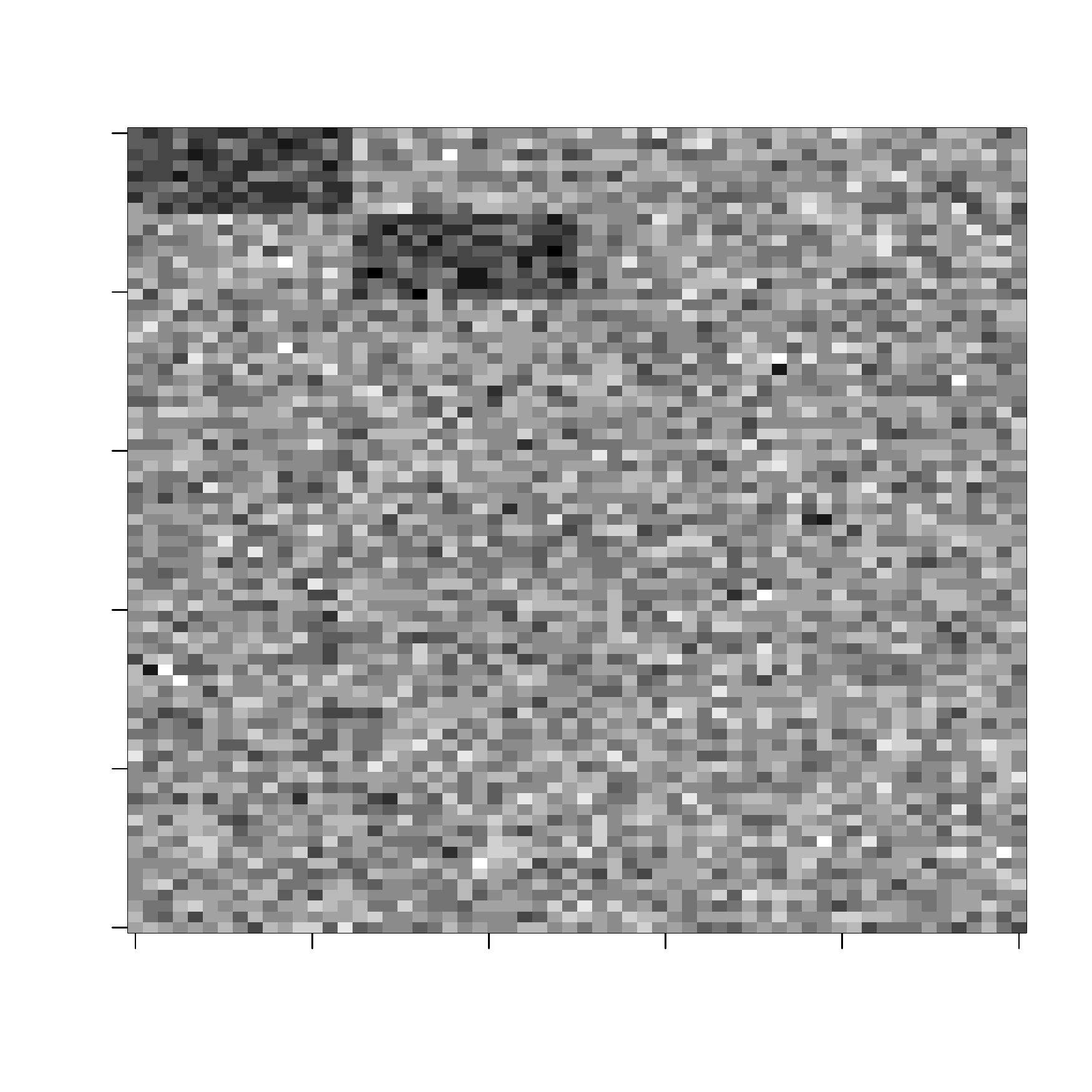}}}
\subfigure[]{\resizebox{55mm}{55mm}{\label{fig:coregulatedexchangeability}\includegraphics{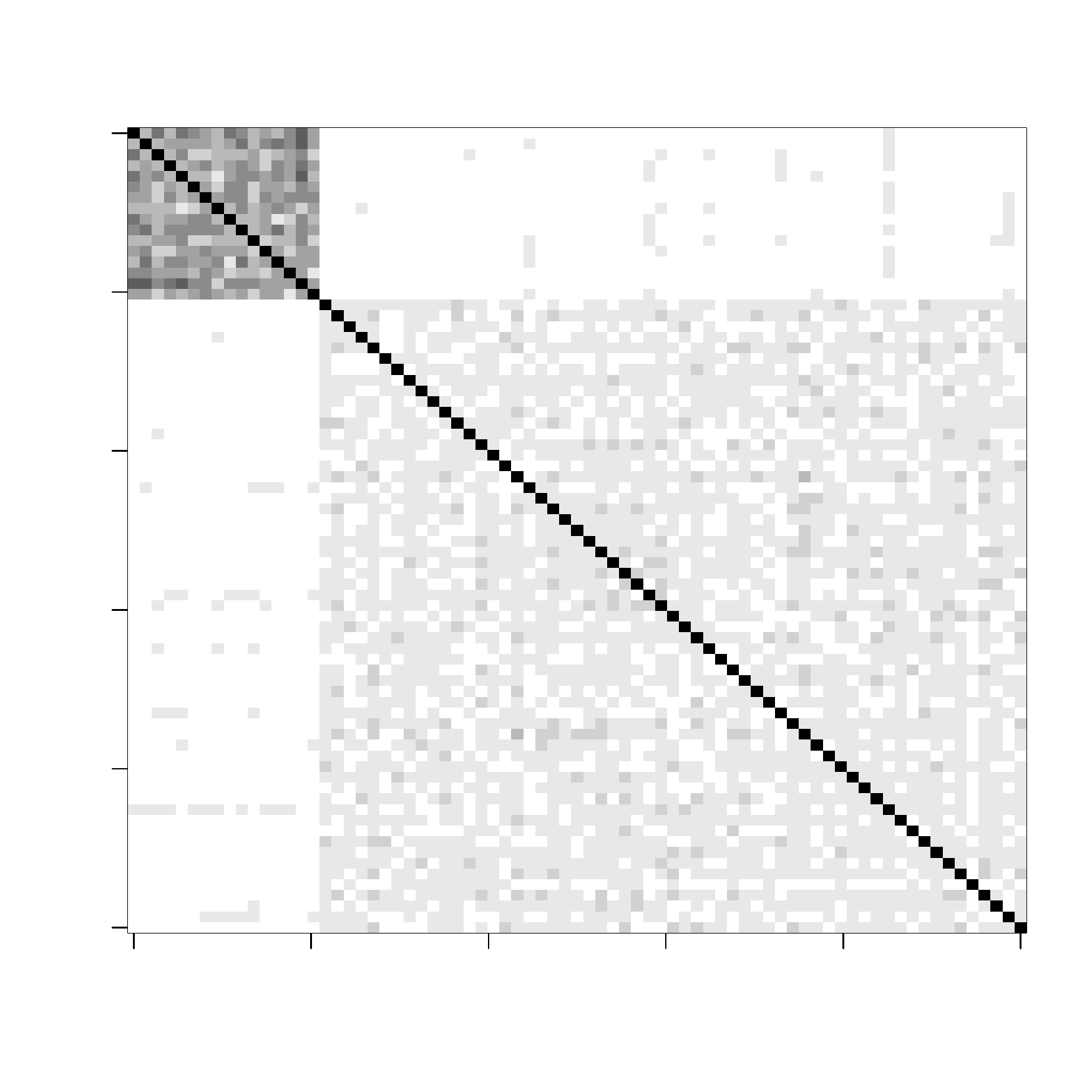}}}
\subfigure[]{\resizebox{55mm}{55mm}{\label{fig:coregulatedcorrelation}\includegraphics{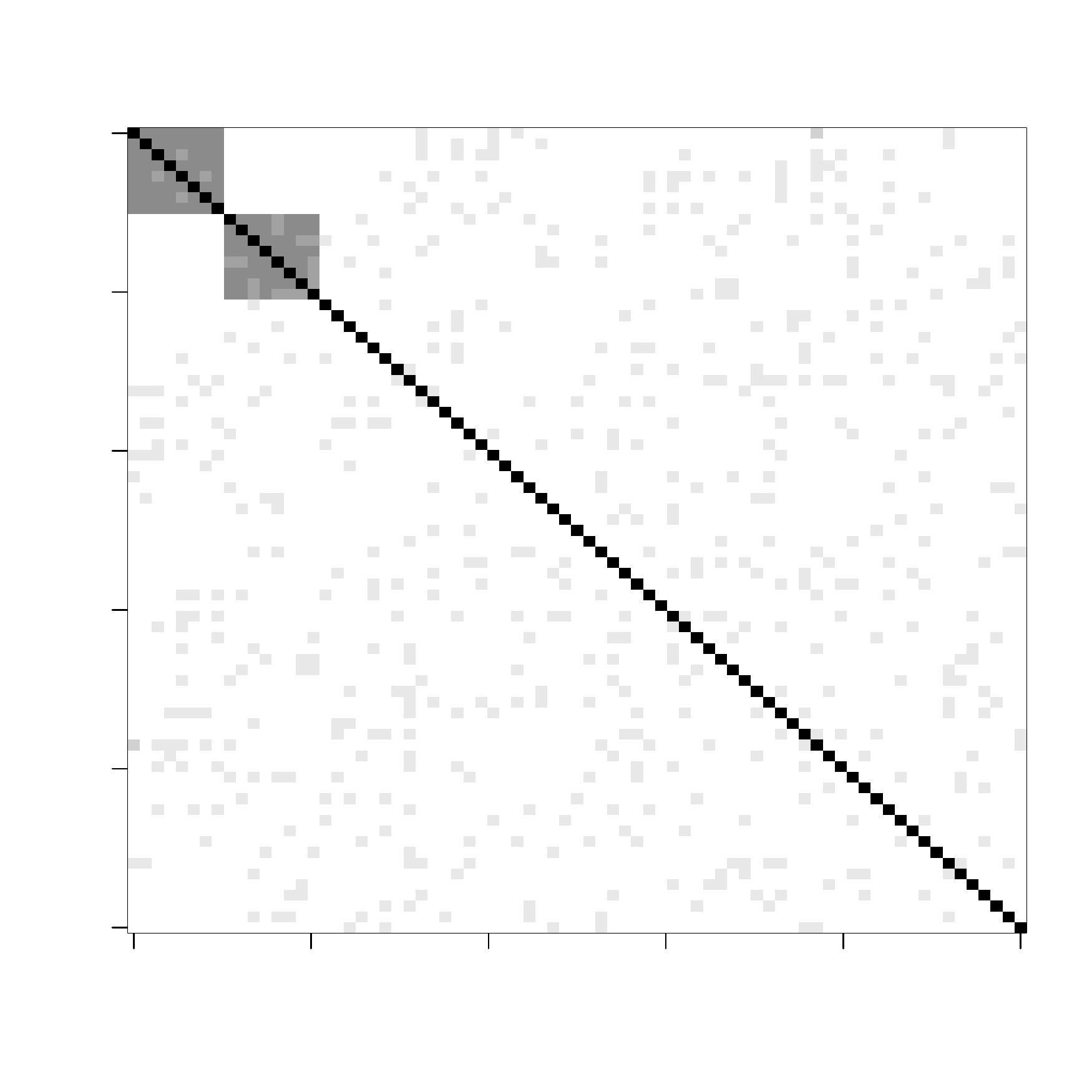}}}
\subfigure[]{\resizebox{55mm}{55mm}{\label{fig:coregulatedcorrexch}\includegraphics{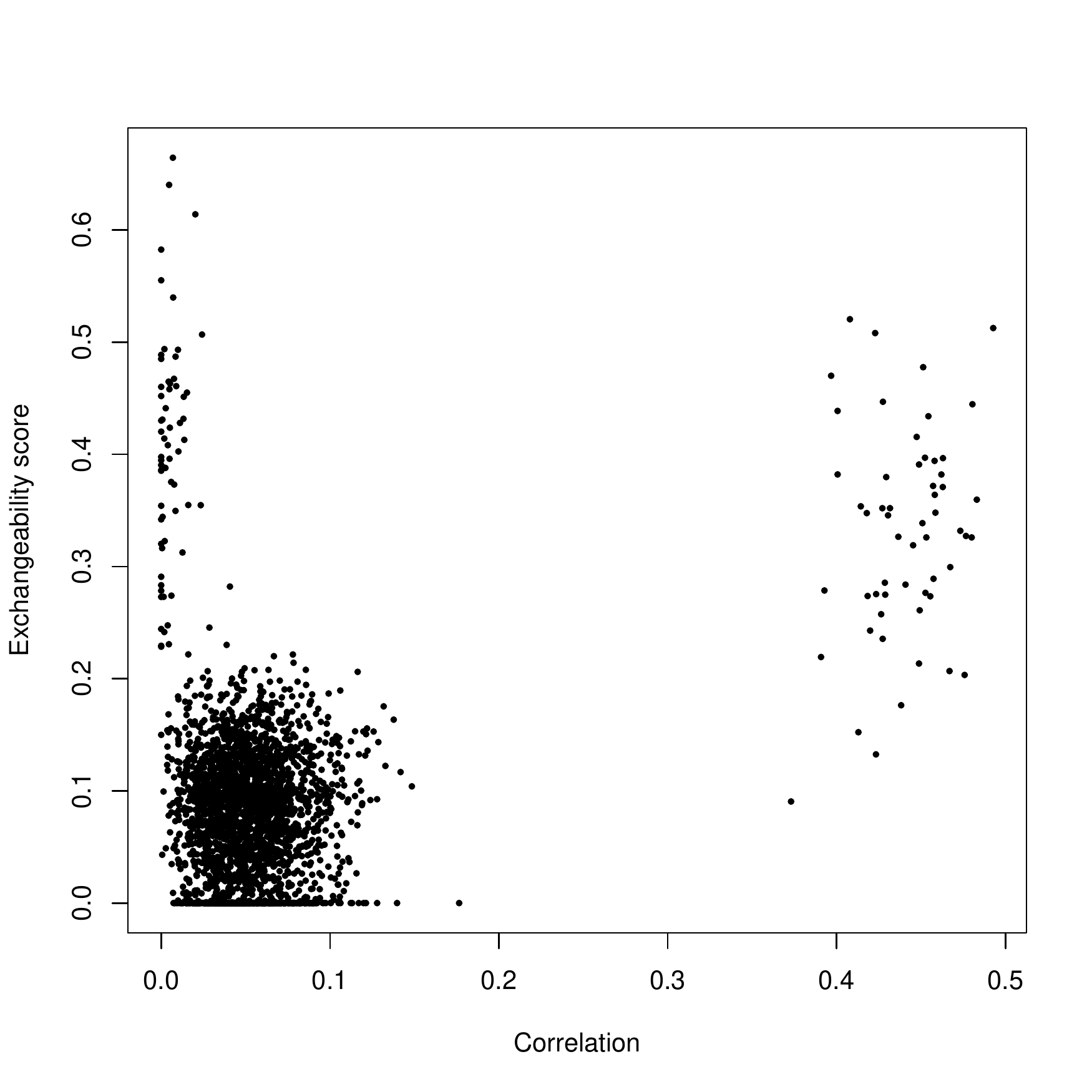}}}
\caption{(a) Structure of the data set in Example~\ref{ex:coregulated}. Each row represents a variable and each column represents a sample. (b) The exchangeability matrix. (c) The positive part of the correlation matrix. (d) The relationship between the positive part of the correlation coefficient and the exchangeability score for the variable pairs. All values are averaged across 10 realizations. }
\label{fig:excoregulated}
\end{center}
\end{figure}
\end{ex}

\section{Methods for list comparison}
In this section, we provide a brief overview of previously proposed methods for list comparison. These comprise both methods for comparing two unordered lists, methods for comparing two ordered lists and methods for comparing one ordered and one unordered list. Reviews can also be found in e.g. \citet{Goeman_Buhlmann_07,Song_Black_08,Ackermann_Strimmer_09,Boulesteix_Slawski_09,Huang_etal_09}. We will also show how a number of the most well-known methods can be formulated within the framework presented in the main article by tuning the selection of $A_\ell$, $V_\ell$, $W_\ell$, $h$ and the similarity or dissimilarity measure on $\mathbb{R}^M\times\mathbb{R}^M$. 

Comparison of gene lists is an essential part of many applications. One such is gene set enrichment analysis, where a (usually short) unordered list (a gene set) is checked for significant enrichment among genes which are highly related to a response. The simplest methods, commonly denoted overrepresentation analysis methods, are based on computing the overlap between the gene set and an unordered set of e.g. differentially expressed genes. The size of the overlap is then checked for significance by comparison to the hypergeometric distribution or its approximation by the binomial or $\chi^2$ distributions \citep{Draghici_etal_03,Hosack_etal_03,Khatri_Draghici_05}. The simplicity of these methods have made them the methods of choice in many software packages and web tools for gene set analysis. In the same spirit, methods based on Venn diagrams have been proposed \citep{Smid_etal_03}, as well as the POG (percentage of overlapping genes) score \citep{EinDor_etal_06,MAQCConsortium_06}. The POG score has recently been extended to take into account correlated molecular changes \citep{Zhang_etal_09} or known functional relationships between the genes \citep{Gong_etal_10}. 

Another approach to gene set enrichment analysis is to first compute a ranking of all the genes from an experiment and then check the genes in the gene set for significant enrichment in the top and/or bottom of the ranking. The most well-known method is Gene Set Enrichment Analysis (GSEA) \citep{Mootha_etal_03,Subramanian_etal_05} where a modified Kolmogorov-Smirnov statistic is used to quantify the enrichment of a gene set. Other methods for combining individual gene statistics into a summary statistic for a gene set, e.g. by simple averaging, have also been pursued \citep[see e.g.][for a discussion and further references]{Ackermann_Strimmer_09}.

To e.g. assess the stability of gene rankings from different studies, several authors have proposed methods for comparing two ordered gene lists. The overlap score proposed by \citet{Yang_etal_06} is one such method, which computes a weighted sum of the overlap of the top-$k$ and/or bottom-$k$ lists for $k=1,\ldots,M$ where $M$ is the number of genes in the ranking. By adjusting the weights, genes in the extreme ends of the lists can have a higher influence on the stability score than the genes in the middle. Viewing the rankings as permutations of $\{1,\ldots,M\}$ (possibly truncated) several methods have been proposed for comparing the corresponding permutations using e.g. Spearman's footrule, Kendall distance or the Canberra metric (or a modified variant for truncated lists) \citep{Fagin_etal_03,Jurman_etal_08,Jurman_etal_10}. \citet{Jurman_etal_08} also account somewhat for the relationship between the genes by including the possibility to define functionally related gene modules, such that the ranking of the variables within such a module does not matter. Other methods for comparing top-$k$ lists taking the ranking into account have been proposed by \citet{Pearson_07} and \citet{Stiglic_Kokol_10}. 

In the rest of this section, we will show how it is possible to formulate many of these methods for gene list comparison in our proposed framework by choosing the basic matrices $A_\ell$, $V_\ell$ and $W_\ell$ as well as the summary function $h$ and the similarity or dissimilarity measure on $\mathbb{R}^M\times\mathbb{R}^M$ suitably. It can be noted that most methods use $V_\ell=I_M$, meaning that associations between the genes in the universal set are not explicitly taken into account. 

First, we note that if we compare two unordered lists $\ell_1$ and $\ell_2$ and choose $V_\ell=W_\ell=I_M$ and $(A_\ell)_{ii}=\chi_\ell(g_i)$ for each of the lists, the similarity score defined by the cosine of the angle between the list vectors is equal to $$s(\ell_1,\ell_2)=\frac{|\ell_1\cap\ell_2|}{\sqrt{|\ell_1||\ell_2|}}.$$This similarity coefficient is the geometric mean of the quantities $\frac{|\ell_1\cap\ell_2|}{|\ell_1|}$ and $\frac{|\ell_1\cap\ell_2|}{|\ell_2|}$ and has been discussed e.g. by \citet{Warrens_08}. 

\subsection{Percentage of overlapping genes-related (POGR)}
\citet{Zhang_etal_09} described a novel metric for quantifying the overlap of two unordered gene sets. Let $k$ denote the number of genes that are shared between the lists $\ell_1$ and $\ell_2$. Then let $O_{r12}$ be the number of genes in $\ell_1$ that are not shared but significantly positively correlated with at least one gene in $\ell_2$. The $POGR$ score is then defined by $$POGR_{12}=\frac{k+O_{r12}}{|\ell_1|}$$and similarly for $POGR_{21}$. We can compute $POGR_{12}$ in our framework by choosing $V_{\ell_1}=W_{\ell_1}=I_M$ and $$(A_{\ell_1})_{ii}=\mathbf{1}\{g_i\in\ell_1\}$$for the first list and $W_{\ell_2}=I_M$, \begin{align*}(A_{\ell_2})_{ii}&=\mathbf{1}\{g_i\in\ell_2\}\\(V_{\ell_2})_{ij}&=\mathbf{1}\{g_i \textrm{ significantly correlated with }g_j\}\end{align*}for the second list. We take $h(x)=\|x\|_\infty$, and then $$POGR_{12}=\frac{l_{\ell_1}\cdot l_{\ell_2}}{\|l_{\ell_1}\|_1}.$$Hence, to compute $POGR_{12}$ we only extend the list vector corresponding to $\ell_2$. Note however that the exchangeability matrix in this case takes both data sets into account, since two genes are significantly correlated only if their expression levels are significantly correlated in both data sets. The reverse score $POGR_{21}$ is calculated by interchanging the roles of $\ell_1$ and $\ell_2$. 

\subsection{Hypergeometric test (Fisher exact test)}
A hypergeometric test, comparing the overlap between two unordered lists $\ell$ and $\ell'$ to what would be expected if they were drawn randomly from the ground set, can be performed by letting $W_\ell=V_\ell=W_{\ell'}=V_{\ell'}=I_M$ and taking $$(A_\ell)_{ii}=\mathbf{1}\{g_i\in\ell\}$$(and similarly for $\ell'$). We take $h(x)=\|x\|_\infty$ and define $$s(\ell,\ell')=l_\ell\cdot l_{\ell'}.$$This gives the size of the overlap between the lists, which is compared to a hypergeometric distribution with parameters $M$, $|\ell|$, $|\ell'|$ to obtain a $p$-value.

\subsection{Gene set enrichment analysis (GSEA)}
GSEA \citep{Mootha_etal_03,Subramanian_etal_05} was developed to estimate the enrichment of the genes within a gene set in a ranking of all variables from an experiment. Hence, the first list ($\ell$) is an ordered list containing all genes in $\mathcal{G}$, and the second list (or gene set) ($\ell '$) is an unordered list with $K$ genes. For $\ell$, we define $$(A_\ell)_{ii}=|r_i|^q,$$
where $r$ is the correlation or ranking metric used to order the genes from the experiment and $q$ is the exponent controlling the weights. 
We take $W_\ell=V_\ell=I_M$. 

For $\ell '$, we choose $$(A_{\ell'})_{ii}=\left\{\begin{array}{lll}|r_i|^q&&\textrm{if }g_i\in\ell '\\-1&&\textrm{if }g_i\not\in\ell '\end{array}\right.$$Note that strictly speaking, this is not a function of the position in the list since all genes in the unordered gene set have the same position. However, as discussed in the main article this may be suitable for gene sets where we wish to incorporate some external information (in this case, information regarding the ranking statistic). We let $$(V_{\ell'})_{ij}=\left\{\begin{array}{lll}1&&\textrm{if }g_i\in\ell ',g_j\in\ell '\textrm{ or }g_i\not\in\ell ',g_j\not\in\ell '\\0&&\textrm{otherwise}.\end{array}\right.$$Finally, we choose $$(W_{\ell'})_{ii}=\left\{\begin{array}{lll}1&&\textrm{if }g_i\in\ell '\\|r_i|^q&&\textrm{if }g_i\not\in\ell '.\end{array}\right.$$Then, we compute the list matrices $G_\ell$ and $G_{\ell'}$ and create the vector representations of the lists by choosing $$h(x)=\sum_{i=1}^Mx_i,$$which means that for the ordered list, $$(l_\ell)_i=|r_i|^q$$ and for the unordered gene set, $$(l_{\ell'})_i=\left\{\begin{array}{lll}\sum_{i;g_i\in\ell '}|r_i|^q&&\textrm{if }g_i\in\ell '\\-(M-K)|r_i|^q&&\textrm{if }g_i\not\in\ell '.\end{array}\right.$$The similarity between the lists (the {\em enrichment score}) is then defined as the maximum deviation from zero of $$\sum_{i=1}^m\frac{(l_\ell)_i}{(l_{\ell'})_i}$$ for $m\in[1,M]$, and the significance of the score is estimated by repeatedly permuting the sample labels and redoing the calculations. 

\subsection{Algebraic comparison of ranked lists}
\citet{Jurman_etal_08} proposed a method for comparing ordered top-$k$ lists, e.g. to estimate the stability of different variable selection methods. Assuming that the two ranked lists to be compared ($\ell$ and $\ell '$ respectively) each contain $K$ genes. We choose $W_\ell=V_\ell=W_{\ell'}=V_{\ell'}=I_M$ and let $$(A_\ell)_{ii}=\left\{\begin{array}{lll}\pi_\ell(i)&&\textrm{if }g_i\in\ell\\K+1&&\textrm{if }g_i\not\in\ell,\end{array}\right.$$with $(A_{\ell'})$ chosen analogously. To create the list vector we use e.g. $h(x)=\min_{1\leq i\leq M}|x_i|$ and we define a dissimilarity measure by $$d(\ell,\ell')=\sum_{i=1}^M\frac{|(l_\ell)_i-(l_{\ell'})_i|}{(l_\ell)_i+(l_{\ell'})_i}.$$The authors also introduce so called {\em feature modules}, consisting of genes known to have similar function. They argue that rank differences within such a module should be less penalized that other differences, and therefore propose to make the distance independent of the ordering within the given modules. In practice, this is obtained by putting the elements of such a module in the same order in all lists by permuting the values in $A_\ell$ and $A_{\ell'}$ corresponding to the genes in the module. 

\subsection{Reciprocal rank-based comparison of ranked lists}
\citet{Pearson_07} proposed a method for comparing ordered top-$k$ lists based on reciprocal ranks. We denote the two ordered lists to be compared  $\ell$ and $\ell'$, and choose $W_\ell=V_\ell=W_{\ell'}=V_{\ell'}=I_M$. We choose $$(A_\ell)_{ii}=\left\{\begin{array}{lll}\frac{1}{\pi_\ell(i)}&&\textrm{if }g_i\in\ell\\\frac{1}{K+1}&&\textrm{if }g_i\not\in\ell\end{array}\right.$$and similarly for $\ell'$. Furthermore, we take e.g. $h(x)=\|x\|_\infty$, and choose $$d(\ell,\ell')=\|l_\ell-l_{\ell'}\|_1.$$

\subsection{Similarity for ordered gene lists}
\citet{Yang_etal_06} presented a method for comparing two rankings $\ell$ and $\ell'$ of all the $M$ genes from an experiment. We let $V_\ell=W_\ell=V_{\ell'}=W_{\ell'}=I_M$, and take $(A_\ell)_{ii}=\pi_\ell(i)$ (and similarly for $\ell'$). Then take e.g. $h(x)=\|x\|_\infty$. This gives \begin{align*}l_\ell&=(\pi_\ell(1),\ldots,\pi_\ell(M))\\l_{\ell'}&=(\pi_{\ell'}(1),\ldots,\pi_{\ell'}(M)).\end{align*}The preliminary similarity score is defined in \citep{Yang_etal_06} as $$S'_\alpha(\ell,\ell')=\sum_{n=1}^Me^{-\alpha n}\left(O_n(\ell,\ell')+O_n(f(\ell),f(\ell'))\right),$$where $\alpha$ is a parameter, $O_n(\ell,\ell')$ is the overlap between the top-$n$ lists of $\ell$ and $\ell'$, and $f(\ell)$ is the list obtained by reversing $\ell$. We have $$\sum_{n=1}^Me^{-\alpha n}O_n(\ell,\ell')=\sum_{n=1}^M\sum_{k=\max((l_\ell)_n,(l_{\ell'})_n)}^Me^{-\alpha k}$$since a gene will contribute to the overlap for all $k$ following its highest position in the two lists. The second term in $S'_\alpha(\ell,\ell')$ can be written accordingly by replacing $(l_\ell)_n$ and $(l_{\ell'})_n$ with $M+1-(l_\ell)_n$ and $M+1-(l_\ell')_n$, respectively. Summation gives $$S'_\alpha(\ell,\ell')=\frac{e^{-\alpha(M+1)}}{1-e^{-\alpha}}\sum_{i=1}^M\left(e^{\alpha(M+1-\max((l_\ell)_i,(l_{\ell'})_i))}+e^{\alpha\min((l_\ell)_i,(l_{\ell'})_i)}-2\right)$$
If it is desirable to allow one list to be the reverse of the other, one can define $$S_\alpha(\ell,\ell')=\max(\beta S'_\alpha(\ell,\ell'),(1-\beta)S'_\alpha(\ell,f(\ell')))$$with $\beta\neq 1$.

\end{document}